\documentclass[12pt,recno]{JHEP3}

\usepackage{graphicx,amsfonts,amssymb,stmaryrd,amsmath,flowdef,longtable}
\title{Defects and Bulk Perturbations of Boundary Landau-Ginzburg Orbifolds}  
\author{Ilka Brunner$^{1}$ and 
Daniel Roggenkamp$^{2}$ \\  
\\
$^{1}$Institut f{\"u}r Theoretische Physik, ETH Z\"urich \\
CH--8093 Z{\"u}rich, Switzerland\\ 
E-mail: \email{brunner@itp.phys.ethz.ch}\\
\\
$^{2}$Department of Physics and Astronomy, Rutgers University\\
Piscataway, NJ 08855-0849, USA\\
E-mail: \email{roggenka@physics.rutgers.edu}}
\abstract{We propose defect lines as a useful tool in the
  study of bulk perturbations of conformal
field theory, in particular in the analysis of the induced renormalisation group 
flows of boundary conditions. As a concrete example we study bulk perturbations
of $N=2$ supersymmetric minimal models. To these perturbations we associate
a special class of defects between the respective UV and IR theories, whose
fusion with boundary conditions indeed reproduces the behaviour of the latter
under the corresponding RG flows.}

\preprint{}

\begin{document}
\section{Introduction}
In string theory, generic non-supersymmetric backgrounds contain tachyons and are therefore unstable.
What happens to such backgrounds under the condensation of these
tachyons is a very interesting question, which is however difficult to
answer in general. 
For a special class of non-supersymmetric orbifolds this has been studied in \cite{Adams:2001sv,Harvey:2001wm,Vafa:2001ra}, where use was made of the fact that the tachyons in these models are localised on 
space-time defects. This allows to apply techniques similar to those employed in the treatment of open string tachyon condensation.
 
From the world sheet point of view, tachyon condensation is decribed by a
perturbation of the conformal field theory associated to the initial background by relevant operators. 
The end point of the induced bulk renormalisation group flow is a new conformal field theory
which describes the vacuum reached after the decay of the original background.

Unfortunately, the analysis of such bulk RG flows is very tedious and in particular for models of interest 
in string theory, not much is known about them. In some cases however there are additional structures which can be used in the analysis. For instance, the non-supersymmetric orbifolds mentioned above, although not being space-time supersymmetric, exhibit $N=2$ world sheet supersymmetry, which gives more controle over the RG flow due to non-renormalisation theorems. 

An interesting question which arises in the context of closed string tachyon condensation
is the fate of the D-branes in the initial theory once the background decays to a new vacuum.
For the non-supersymmetric orbifolds this has been addressed in \cite{Martinec:2002wg,Moore:2004yt,Moore:2005wp}. Since the tachyon condensation
in these examples partially resolves the orbifold singularity, there are fewer D-brane charges
available after the condensation, and some of the D-branes have to decouple from the theory.

From a world sheet point of view, the effect of closed string tachyon condensation on D-branes
is described by perturbations with relevant bulk fields in the presence of world sheet boundaries.
Such perturbations induce flows in both the bulk as well as the boundary sectors of the theory,
which makes them even more tedious to analyse. 

In this paper, we propose a new approach to the regularisation and
renormalisation of bulk perturbations in the presence of boundaries.
As explained in some detail in Section 2, we decouple bulk and boundary
flows. We first perform the bulk flow and, in a second step, treat the effect
on the boundary sectors, which then amounts to merging a world sheet defect
line with a boundary.

Defects lines are one-dimensional interfaces
which separate two possibly different conformal field theories 
(see \eg \cite{Petkova:2000ip,Bachas:2001vj,Frohlich:2004ef,Bachas:2004sy,Frohlich:2006ch,Quella:2006de,Alekseev:2007in,Fuchs:2007tx,Runkel:2007wd}).
A special type of such defects, so called {\it topological defects} can be shifted on the world sheet
and in particular can be moved smoothly on top of other
defects resulting in new {\it fused} defects. Likewise, they can be brought to
world sheet boundaries transforming the original boundary conditions to different ones.
Generic, non-topological defects on the other hand, cannot be moved on the world sheet without changing
correlation functions, and bringing them close to world sheet boundaries (or to other such defects) results in singularities.

The defects which emerge in our treatment of bulk perturbations of boundary conformal field theories
are defects between the UV and IR CFTs of the corresponding RG flows. As such they are non-topological in general, and hence their fusion with world sheet boundaries is singular. This is indeed expected. 
Encoding the effect of the bulk perturbation near the boundary, the process of merging the defect with the
boundary has to be regularised just as the bulk perturbation near the boundary has to be.

To avoid dealing with this regularisation, we consider supersymmetric flows between $N=2$ superconformal field theories here. The corresponding defects are supersymmetric and are in particular
compatible with a topological twist of the theory. 
In the twisted theory, they become topological
and can therefore be merged with world sheet boundaries without the need of regularisation, determining in this way to what boundary condition a given one flows under the bulk perturbation.

The concrete examples we study here are RG flows between orbifolds of $N=2$ superconformal minimal models generated by twist field perturbations. These models have an alternative description as Landau-Ginzburg orbifolds which lends itself easily to the construction
of supersymmetric defects and the analysis of their properties as \eg fusion. We generalise
the formalism developed in \cite{Khovanov:2004bc,Kapustin:2004df,Brunner:2007qu} to deal with B-type defects in Landau-Ginzburg models to the case of Landau-Ginzburg orbifolds, and use it to construct the defects describing the twist field perturbations of minimal model orbifolds. 
Indeed, 
these flows are very similar to the flows of non-superymmetric orbifolds mentioned above, and the methods
we describe here easily generalise to perturbations of orbifolds $\CC/\ZZ_d$.

The use of defects to describe the effect of bulk flows on boundary conditions is non-perturbative in nature.
After all, it involves a defect between UV and IR conformal field theories. Thus, it is in general difficult to 
find the defect describing a particular perturbation.
In the examples at hand however, one can use mirror symmetry to relate perturbations and defects.
The mirrors of the minimal model orbifolds are the unorbifolded minimal models, which have a Landau-Ginzburg description. The twist fields perturbing the minimal model orbifolds are mapped under mirror symmetry to monomials in the chiral superfield deforming the superpotential of the Landau-Ginzburg model. The effect of such deformations on A-branes can easily be studied and compared to the fusion of defects
in the Landau-Ginzburg orbifolds.

This paper is organised as follows.
In Section \ref{bulkflows} we describe in some more detail how the effect of bulk perturbations on boundary conditions
is captured by the fusion with a defect. In Section \ref{setsum} we introduce the
concrete examples (perturbations of orbifolds of $N=2$ superconformal field theories), in which we apply this method, and at the same time outline our strategy and summarise the results obtained in the following sections.
Section \ref{deflgorb} is devoted to a general discussion of B-type defects in Landau-Ginzburg orbifolds and their description 
in terms of equivariant matrix factorisations.
In Section \ref{flowdefects} we use this formalism to construct a special class of defects between orbifolds of $N=2$ minimal models,
which we propose to be the defects arising in RG flows between these models. We also analyse their properties, in particular their fusion with each other and with B-type boundary conditions. 
In Section \ref{comparison} we compare this fusion with the behaviour of A-type D-branes under 
the corresponding flows on the mirror side, which can be studied rather explicitly.
Section \ref{orbsec} contains some comments on the description of RG flows for
non-supersymmetric orbifolds $\BC/\ZZ_d$. We close with some open problems in Section \ref{discussion}.
\section{Bulk Flows and Defects}\label{bulkflows}
The topic of this article is the behaviour of D-branes or conformal boundary condition
under relevant bulk perturbations. This subject has been addressed in the literature
before, using the Thermodynamic Bethe Ansatz \cite{Dorey:1997yg} 
the truncated conformal space method \cite{Dorey:2000eh} or by analysing the 
RG flow equations for bulk and boundary couplings \cite{Fredenhagen:2006dn,Keller:2007nd}.
The new idea put forward in this article is to use defects to describe the effect of bulk perturbations of conformal field theories on conformal boundary conditions.

Conformal field theories can be perturbed by adding terms
\beq
\Delta S=\sum_i\lambda^i \int_\Sigma{\rm d}^2z\varphi_i(z,\bar z)
\eeq
to the action. Here $\lambda^i$ are coupling constants, and $\varphi_i$ are marginal or relevant
fields which are integrated over the world sheet surface $\Sigma$. Perturbed correlation functions
are then obtained from those of the unperturbed theory by
\beq\label{perturbation}
\langle\ldots\rangle_{\lambda^i}=\langle \ldots e^{\Delta S}\rangle_{\lambda^i=0}\,.
\eeq
Obviously, expressions like this have to be regularised for instance by means of a cutoff restricting the
integration domain of the perturbations away from any other field insertion. The renormalisation group flow
then drives the system from the UV to the IR fixed point (if existent) of the perturbation, which is another conformal field theory. 
In the special case where the operator is marginal, the theory remains conformal for all values of the
coupling constants.

Instead of a perturbation on the entire surface, one can also consider perturbations which are
restricted to a domain $U\subset\Sigma$, as indicated in figure \ref{flowfigure}.
\FIGURE[ht]{
\includegraphics[width=9cm]{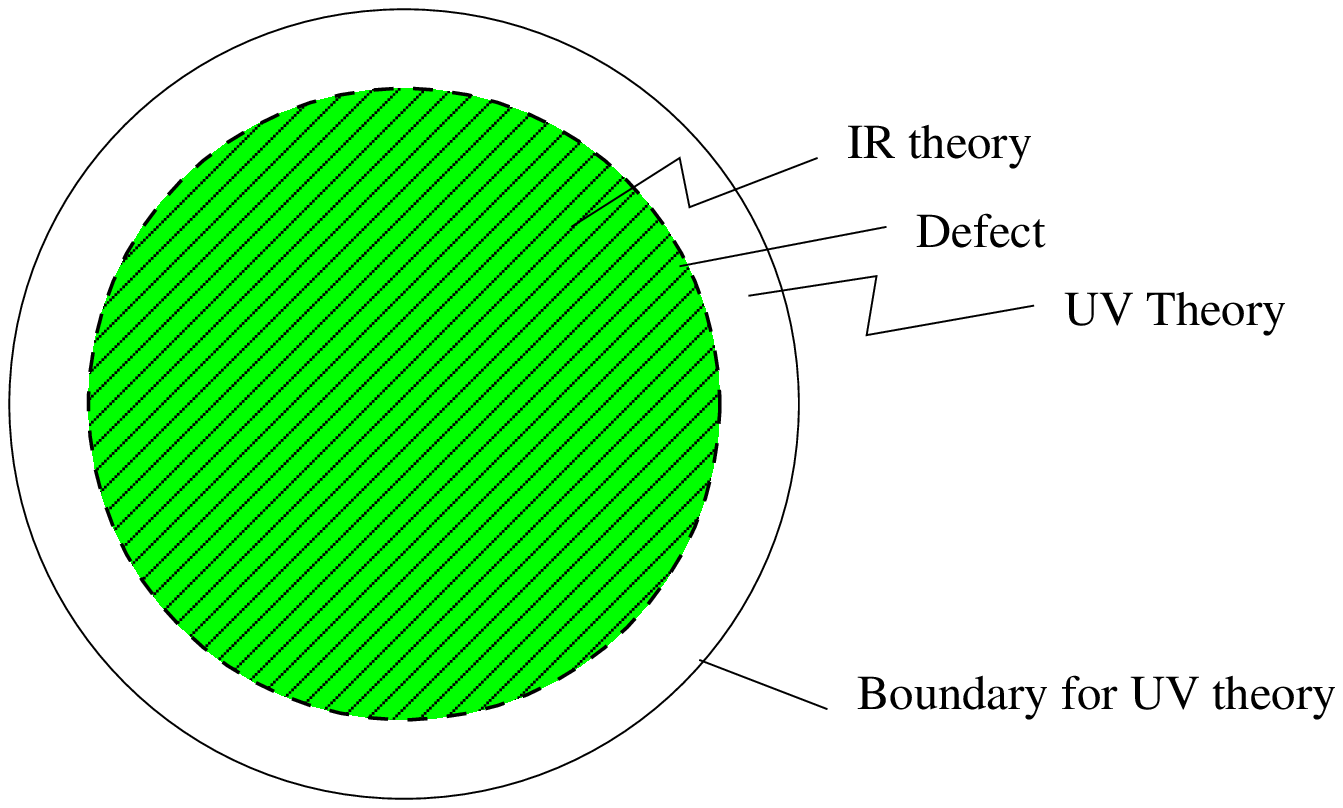}
\caption{Perturbation restricted to a domain $U$ (shaded). UV
and IR theory are separated by a defect line.\label{flowfigure}}
}
In the same way as before, one obtains perturbed correlation functions and a renormalisation
group flow. Since local properties outside $U$ are not affected by the perturbation, the correlation functions
at the endpoint of the flow describe the situation of the original UV conformal field theory on $\Sigma-U$ and the
IR theory on $U$, which because of conformal invariance are separated by a conformal defect line on the boundary 
$\partial U$. In this way, perturbations give rise to conformal defects separating UV and IR theories of the corresponding 
renormalisation group flows. (Similarly, perturbations with exactly marginal operators give rise to defects between
the unperturbed and the perturbed theory.)

This relation between bulk RG flows and defects is particularly useful in the study of bulk perturbations of
conformal field theories on surfaces with boundary\footnote{See
\cite{Fredenhagen:2006dn,Baumgartl:2007an,Green:2007wr} for recent discussions
of bulk induced boundary flows using other methods.}. 
Apart from the bulk RG flow such perturbations in general also
induce RG flows in the boundary sectors, and UV boundary conditions flow to 
boundary conditions of the IR theory.
It is a very interesting question, to which IR boundary condition a given boundary condition in the UV flows, 
or to formulate it in string theory terminology, what happens to a D-brane under closed string tachyon condensation.

If in the UV one starts with a conformal field theory defined on a surface $\Sigma$ with boundary
and a conformal boundary condition along $\partial\Sigma$, then, besides the regularisation already present in the bulk case, 
one also
has to restrict the domains of the integrals \eq{perturbation} away from the boundary. This is due to a 
non-trivial singular bulk-boundary operator product expansion. Thus, in this situation there are two independent regularisation parameters, one of 
which parametrises the RG flow in the bulk, whereas the other one parametrises the induced flow in the boundary sectors.
While these two flows are often treated simultaneously, we propose to perform the bulk flow first, while keeping the
boundary regularisation fixed. This is nothing but a bulk flow on the subdomain 
\beq
U_\epsilon:=\{z\in\Sigma\,|\,{\rm dist}(z,\partial\Sigma)\geq \epsilon\}\subset\Sigma
\eeq
of all points on $\Sigma$ whose distance from the boundary $\partial\Sigma$
is bigger than the boundary regularisation parameter $\epsilon$.
Hence, the endpoint of the pure bulk flow with fixed boundary regularisation parameter
 is the IR theory on $U_\epsilon$ separated by
a conformal defect line from the UV theory defined on the neighbourhood $\Sigma-U_\epsilon$ of the boundary. 

The second step, namely the flow in the boundary sector is then described by letting $\epsilon$ go to zero, 
and in that way bringing the defect towards the boundary. This procedure produces
out of the UV boundary condition a boundary condition of the IR theory.
However, a priori this is a singular process,  because the defect is a non-topological defect in general. 
That is not surprising. After all, the singularities appearing in the correlation functions when the perturbing fields
$\varphi_i$ come close to the boundary have not been cancelled by counterterms, because we have not performed the boundary RG flow.
In fact, all the singularities arising at the boundary due to the entire bulk flow are encoded in the defect, and the
process of taking the defect to the boundary has to be regularised in an appropriate way to obtain the induced flow
in the boundary sectors.

Note that the approach we propose to describe the effect of bulk perturbation on boundary conditions is non-perturbative
in the bulk coupling constants. Namely, the 
defect we associate to a bulk perturbation connects directly
UV and IR theories of the corresponding RG flow. This obviously is an advantage, at least in the case where one can make sense of
the procedure of bringing the defect close to the boundary. On the other hand, it is often not obvious how to connect
perturbative and non-perturbative descriptions, {\it i.e.} how to find the defect associated to a particular bulk perturbation.

The structure of fusion of non-topological defects with boundaries is a very interesting subject, and has been considered for the case of the free boson
in \cite{Bachas}. 
In this paper however
we will avoid all intricacies arising in this context
by considering $N=2$ superconformal field
theories.
These theories can be topologically twisted, which in particular makes all defects preserving the appropriate supersymmetries topological. That means that they can be shifted on the surface without changing correlation functions, and in particular without giving rise to singularities when brought near boundaries or other defects.
In this way there is a well defined fusion of supersymmetry preserving defects with supersymmetric boundary conditions or defects.

Our purpose in the following is to identify the supersymmetric defects associated to supersymmetry preserving
perturbations of $N=2$ superconformal field theories in the way described above.
Their fusion with supersymmetric boundary conditions then describes to which 
boundary conditions the latter flow in the IR.

There are two kinds of supersymmetry preserving perturbations one can consider in  $N=2$ supersymmetric theories.
Firstly there are pertrubations of the bulk action by chiral superfields $\Phi$ integrated over the chiral half of superspace\footnote{Conventions on superspace are taken from \cite{Hori:2000ic}.}
\beq\label{bper}
\Delta S_c = \int_\Sigma d^2x \, d\theta^- d\theta^+ \left. \Phi
\right|_{\bar\theta^\pm = 0}  + c.c.\,.
\eeq
By construction, on surfaces without boundaries this perturbation leaves supersymmetry unbroken. On surfaces with boundaries however
this is no longer the case, and the variation of the action gives rise to a boundary term \cite{Hori:2000ck,Hori:2000ic}.
In case of an A-type boundary, this term can be compensated by the supersymmetry variation of a boundary integral of the form
\beq\label{Bdef}
{\cal B} = i \int_{\partial \Sigma } ds (\phi - \bar\phi)
\eeq
which can be added to the action in order to preserve supersymmetry.
If the boundary is of B-type, the boundary term cannot in general be cancelled in this manner, and 
perturbations of type \eq{bper} induce non-supersymmetric boundary flows.

The second type of supersymmetric bulk perturbation is given by integrals 
\beq\label{twbper}
\Delta S_t = \int_\Sigma d^2x \, d\bar\theta^- d\theta^+ \left. \Psi 
\right|_{\bar\theta^+=\theta^- = 0} +c.c.
\eeq
of twisted chiral superfields $\Psi$.
In agreement with mirror symmetry, the boundary terms resulting from varying the action can be cancelled for
B-type boundaries, but not for A-type ones.

Here, we are interested in bulk flows which also preserve supersymmetry in the boundary sectors.
Thus, we can consider either chiral perturbation in the presence of A-type boundary conditions or
twisted chiral perturbations in the presence of B-type boundary conditions\footnote{In the context of 
non-linear sigma models with K\"ahler target space, this corresponds to perturbations of the complex
structure in the presence of A-branes or perturbations of the K\"ahler structure in the presence of B-branes.}.
As discussed above, performing the bulk RG flow while keeping the boundary regularisation parameter $\epsilon$ 
fixed, one obtains the IR theory on a domain $U_\epsilon\subset\Sigma$ and the IR theory on the neighbourhood 
$\Sigma-U_\epsilon$ of the boundary $\partial\Sigma$, which are separated by a defect on $\partial U_\epsilon$. 

Turning the arguments above around shows 
that perturbations with chiral superfields on a domain $U$
give rise to A-type defects, whereas perturbations with twisted chiral superfields give rise to
B-type defects. Therefore, to identify what happens to the respective boundary conditions under a bulk flow,
one first has to identify the corresponding A- or B-type defect 
and then analyse its fusion with the boundary condition.
As alluded to above, the latter requires regularisation, but one can use supersymmetry to avoid dealing with it explicitly. 
Since the flow is supersymmetric all
along, one can consider it in the topologically twisted theory, in which fusion of defects and boundary conditions is non-singular. 
This permits to identify the flows of topological boundary conditions, which in the situation we will consider here
is enough to conclude the flows of the correspoding supersymmetric boundary conditions in the full conformal field theories.
\section{Setup and Outline}\label{setsum}
In the following we will apply and test the method outlined in Section \ref{bulkflows} in the case of orbifolds ${\cal M}_{d-2}/\ZZ_d$
of $N=2$ superconformal minimal models. 
These are well studied rational conformal field theories at central charge $c=3(1-{2\over d})$. (A few details about them are
collected in Appendix \ref{cftapp}.) We are interested in perturbations which preserve supersymmetry.
As discussed in Section \ref{bulkflows} there are two types of such perturbations, chiral and twisted chiral ones. 
The chiral ones are generated by $(c,c)$-chiral primary fields\footnote{We are only interested in unitary flows, 
so that we always perturb with a conjugation invariant operator.}
and the twisted chiral ones by
$(a,c)$-chiral primary fields. The corresponding perturbations are marginal or relevant if these fields have conformal
weights $\leq{1\over 2}$. 
As it turns out, in ${\cal M}_{d-2}/\ZZ_d$ the $(c,c)$-chiral ring is trivial.
But there are $(a,c)$-fields, which can be used to perturb the theory. More precisely, for every $i=0,\ldots d-1$
there is an $(a,c)$-field $\Psi_i$ of conformal weights $h_i=\overline{h}_i={i\over 2d}<{1\over 2}$, which can be obtained by spectral flow from the unique Ramond ground state in the $i$th twisted sector of the theory. $\Psi_0$ is the identity field, but all the other ones generate relevant perturbations, which drive the theory in the IR to another minimal model orbifold with smaller $d$ however.

To understand these renormalisation group flows, the Landau-Ginzburg
realisation of the involved models is very useful. The minimal model ${\cal M}_{d-2}$ can be obtained as IR limit of an $N=2$ Landau-Ginzburg model with a single chiral superfield $X$ and superpotential $W=X^d$. The orbifold group $\ZZ_d$ acts in the Landau-Ginzburg model by multiplication of $X$ by $d$th roots of unity, so that the orbifold model ${\cal M}_{d-2}/\ZZ_d$ 
can be obtained as IR limit of the corresponding Landau-Ginzburg orbifold. 

The RG flows can now be formulated in the framework of gauged linear sigma models, analogous to the flows between affine
orbifold models $\CC/\ZZ_d$ considered in \cite{Vafa:2001ra}. We will not describe this approach here. Instead, we will
study the perturbation in the mirror representation. The mirror of a minimal model orbifold ${\cal M}_{d-2}/\ZZ_d$ is just
the minimal model ${\cal M}_{d-2}$ itself (see Appendix \ref{cftapp}), and the mirror of the $(a,c)$-fields $\Psi_i$ are the fields
corresponding to the monomials $X^i$ in the superfield $X$ of the associated Landau-Ginzburg model. Thus, on the mirror side,
perturbations generated by the $\Psi_i$ are described by a Landau-Ginzburg model with superpotential $W=X^d$ deformed by
lower order terms. Not being homogeneous anymore, the deformed superpotential effectively flows under the renormalisation group
due to field redefinitions (see \eg \cite{Vafa:2001ra}). It flows to a homogeneous superpotential corresponding to another conformal field theory. 

As an example consider the perturbation of the orbifold model by the field $\Psi_{d\p}$. 
On the mirror side this corresponds to deforming the superpotential $W=X^d$ to $W=X^d+\lambda X^{d\p}$. The RG flow has the effect of scaling the superpotential as $W\mapsto\Lambda^{-1}W$. Accompanied by a field redefinition $X\mapsto \Lambda^{1\over d} X$, this yields a running coupling constant $\lambda(\Lambda)=\lambda_0\Lambda^{d\p-d\over d}$. In the UV ($\Lambda\rightarrow\infty$) the coupling goes to zero, whereas in the IR ($\Lambda\rightarrow 0$), the coupling diverges. Hence, this describes a flow between the
Landau-Ginzburg models with superpotentials $W=X^d$ in the UV and the one with $W=X^{d\p}$ in the IR. Therefore, the relevant operator $\Psi_{d\p}$ induces
an RG flow between the orbifolds ${\cal M}_{d-2}/\ZZ_d$ and ${\cal M}_{d\p-2}/\ZZ_{d\p}$.  

Our aim is to study flows of this type, and in particular their effect on boundary conditions 
using defects between the minimal model orbifolds in UV and IR. 
Such defects cannot be topological, because the two conformal field theories are connected by a relevant
flow and hence have different central charge. The construction of non-topological defects between conformal field
theories is difficult, but since the flows preserve supersymmetry also the defects have to be supersymmetric. More precisely, being generated by twisted chiral fields, the flows give rise to B-type defects (\cf Section \ref{bulkflows}). Thus we can make use of
a nice description of B-type defects between Landau-Ginzburg models in terms of matrix factorisations of the the difference 
of the respective superpotentials \cite{Brunner:2007qu}. This formalism not only lends itself easily to the construction of defects, but also to the analysis of their fusion and their fusion with B-type boundary conditions, which are also represented in terms of matrix factorisations \cite{Kontsevich,Kapustin:2002bi,Orlov,Brunner:2003dc}. 

To be applicable to the study of defects between minimal model orbifolds, 
it has to be generalised to orbifolds of Landau-Ginzburg models.
Similarly to B-type boundary conditions \cite{Ashok:2004zb,Hori:2004ja}, also B-type defects between 
Landau-Ginzburg orbifolds are described by matrix factorisations which are equivariant with respect to the action of 
the orbifold group, and properties like fusion generalise in a straight forward manner. This is discussed in Section \ref{deflgorb}.

With this formalism at hand, in Section \ref{flowdefects} we construct a class of defects between minimal model orbifolds
${\cal M}_{d-2}/\ZZ_d$ with different $d$ which we propose to arise in renormalisation group flows between these theories.
This class of defects closes under fusion, \ie the fusion of such a defect
between minimal model orbifolds ${\cal M}_{d-2}/\ZZ_d$ and ${\cal M}_{d\p-2}/\ZZ_{d\p}$ and 
one between ${\cal M}_{d\p-2}/\ZZ_{d\p}$ and ${\cal M}_{d\p\p-2}/\ZZ_{d\p\p}$
is a defect between ${\cal M}_{d-2}/\ZZ_d$ and ${\cal M}_{d\p\p-2}/\ZZ_{d\p\p}$ of the same type. 
This fusion of defects indeed corresponds to the concatenation
of renormalisation group flows. 

As alluded to above, the flows we are interested in can be very explicitly studied
in the mirror Landau-Ginzburg models, where they are just given by deformations of the superpotential. 
In particular, this can be used to investigate what happens to B-type boundary conditions in ${\cal M}_{d-2}/\ZZ_d$ under the flows. 
Namely, the corresponding mirror A-type boundary conditions have a very
nice geometric interpretation as Lefshetz pencils of the superpotential $W$ \cite{Hori:2000ck}, whose behaviour under
deformations of $W$ can be determined explicitly. In Section \ref{comparison} we compare these flows to the fusion
of our special defects with B-type boundary conditions calculated in Section \ref{flowdefects} 
and find complete agreement. This provides strong evidence for our claim
that the special defects indeed are the ones which arise in the renormalisation
group flows between different minimal model orbifolds.

As alluded to above, the flows between minimal model orbifolds ${\cal M}_{d-2}/\ZZ_d\mapsto{\cal M}_{d\p-2}/\ZZ_{d\p}$ studied here are very similar to flows between affine orbifold models $\CC/\ZZ_d\mapsto\CC/\ZZ_{d\p}$. In Section \ref{orbsec} we 
argue that our special defects between ${\cal M}_{d-2}/\ZZ_d$ have counterparts in the affine orbifolds $\CC/\ZZ_d$, which describe
the corresponding flows there. 
\section{B-type Defects in Landau-Ginzburg Orbifolds}\label{deflgorb}
Although the method described in Section \ref{bulkflows} above can be applied to study supersymmetric bulk flows 
of any theory with $N=2$ supersymmetry,
we will restrict our further discussion to $(a,c)$-perturbations of
Landau-Ginzburg orbifold models. These can be described by means of B-type defects,
and they remain supersymmetric on surfaces with boundary as long as B-type boundary conditions
are imposed. 
B-type boundary 
conditions for Landau-Ginzburg models of chiral superfields $X_i$ 
can be represented by matrix factorisations of the superpotential $W(X_i)$
\cite{Kontsevich,Kapustin:2002bi,Orlov,Brunner:2003dc}. 

In \cite{Brunner:2007qu}, see
\cite{Khovanov:2004bc,Kapustin:2004df,Kov-Roz} for earlier work in a
different context, it was shown that likewise
B-type defects between two Landau-Ginzburg models with
chiral superfields $X_i$ and $Y_i$ 
and superpotentials $W_1(X_i)$ and $W_2(Y_i)$ respectively can be described by means of matrix factorisations
\beqn\label{matrixfactorisations}
&&P:\quad P_1\overset{p_1}{\underset{p_0}{\rightleftarrows}} P_0\,,\\
&&\quad p_1p_0=(W_1(X_i)-W_2(Y_i))\id_{P_0}\,,\quad 
p_0p_1=(W_1(X_i)-W_2(Y_i))\id_{P_1}\,.\nonumber
\eeqn
of the difference $W_1(X_i)-W_2(Y_i)$ of superpotentials. Here $p_i$ are homomorphisms between the free
$S=\CC[X_i,Y_i]$-modules $P_i$. 

In this section, we will extend the methods developed for Landau-Ginzburg
defects to the case of Landau-Ginzburg orbifolds, in which defects can be 
represented by equivariant matrix factorisations. In particular 
fusion of defects and of
defects with boundary conditions will be formulated in this formalism.

\subsection{Defects and Equivariant Matrix Factorisations}
In the same way as for boundary conditions
\cite{Ashok:2004zb,Hori:2004ja} the matrix factorisation formalism for defects
can be generalised to orbifolds of Landau-Ginzburg models. 
Namely, let $\Gamma_1$ and $\Gamma_2$ be orbifold groups of the
respective LG models, 
\ie $\Gamma_1$ acts on $\CC[X_i]$ and $\Gamma_2$ on $\CC[Y_i]$ in a way compatible 
with multiplication in these rings, such that $W_1(X_i)$ and $W_2(Y_i)$ are invariant. 

 A defect between the respective Landau-Ginzburg orbifolds is then given by a $\Gamma:=\Gamma_1\times\Gamma_2$-equivariant matrix factorisation of $W_1(X_i)-W_2(Y_i)$. The latter is a matrix factorisation \eq{matrixfactorisations} together with
 representations $\rho_i$ of $\Gamma$ on the modules $P_i$ which are compatible with the
 $S$-module structure and with the maps $p_i$. That means that the $\Gamma$-action on the $P_i$ defined by $\rho_i$
 satisfies
 \beq
 \rho_i(\gamma)(s\cdot p)=\rho(\gamma)(s)\cdot\rho_i(\gamma)(p)\,,\quad \forall \gamma\in\Gamma\,,\;s\in S\,,\;p\in P_i\,,
 \eeq
 where $\rho$ denotes the action of $\Gamma$ on $S$, and that furthermore the maps $p_i$ commute with the actions $\rho_i$:
 \beq
 \rho_0(\gamma)p_1=p_1\rho_1(\gamma)\,,\quad
 \rho_1(\gamma)p_0=p_0\rho_0(\gamma)\,,\quad\forall \gamma\in\Gamma\,.
 \eeq
 More details on equivariant matrix factorisations can be found \eg in \cite{Ashok:2004zb,Hori:2004ja}.

In the cases we are interested in here, the orbifold groups are commutative.
In particular, their action give the polynomial rings $S$
the structure of graded rings, and the representations $\rho_i$ turn the $P_i$ into graded $S$-modules. Compatibility furthermore ensures that the maps $p_i$ respect the grading, \ie they are graded of degree $0$.
Matrix factorisations which are equivariant with respect to an abelian group action are therefore sometimes referred to as graded matrix factorisations. 

Note that not all matrix factorisations $P$ admit such representations $\rho_i$. Since the $P_i$ are free $S$-modules,
compatibility with the $S$-action is easily achieved, but the compatibility with the homomorphisms $p_i$ is a non-trivial constraint. 
However, there is a standard procedure to construct from any matrix factorisation $P$ a $\Gamma$-equivariant one by the 
orbifold construction, known for instance from the construction of boundary conditions in general 
orbifold theories from boundary conditions in the underlying non-orbifolded models. Given any matrix factorisation $P$,
one considers the normal
subgroup $\Gamma\p\subset\Gamma$, which stabilises the matrix\footnote{The $p_i$ are matrices with entries in $S$, on which $\Gamma$ acts.} $p_1$, hence also $p_0$, up to change of basis. 
Then one chooses a representation of $\Gamma\p$ on $P$, and extends it to a $\Gamma$-representation of the matrix factorisation given by the sum of the $\Gamma/\Gamma\p$-orbit\footnote{The group $\Gamma/\Gamma\p$ acts on the
matrix factorisations stabilised by $\Gamma\p$.} of $P$
\beq
\wt{p}_i:=\bigoplus_{\gamma\in\Gamma/\Gamma\p}\gamma(p_i)\,,\qquad \wt P_i=\CC[\Gamma/\Gamma\p]\otimes P_i\,.
\eeq
This obviously defines an equivariant matrix factorisation of $W_1(X_i)-W_2(Y_i)$. 

Given two $\Gamma$-equivariant matrix factorisations, the compatibility properties of the representations $\rho_i$
ensure that the $\Gamma$-action lifts to an action on the corresponding BRST-cohomology groups of the matrix factorisations. The BRST-cohomology groups in the orbifold theories are then given by the $\Gamma$-invariant subgroups of the BRST-cohomology groups of the underlying matrix factorisations:
\beq
\HH_{\rm orb}^*(P,Q)=\left(\HH^*(P,Q)\right)^\Gamma\,.
\eeq
\subsection{Fusion}
The most important property of defects which we will use is their fusion. The fusion of B-type defects in Landau-Ginzburg models has been discussed in \cite{Brunner:2007qu}. Let us consider the situation of three LG models with chiral superfields $X_i$, $Y_i$, $Z_i$ and superpotentials $W_1(X_i)$, $W_2(Y_i)$ and $W_3(Z_i)$ respectively. A B-type defect between the first two of these models can be fused with a B-type defect of the last two giving rise to a B-type defect between the first and the last of these models. On the level of matrix factorisation this can be described as follows. The first of these defects can be represented by a matrix factorisation $P$ of $W_1(X_i)-W_2(Y_i)$, whereas the second one is described by a matrix factorisation $Q$ of $W_2(Y_i)-W_3(Z_i)$. The defect obtained by fusion of the two can now be represented by the matrix factorisation $P*Q$, which is defined to be the tensor product matrix factorisation $P\otimes Q$
\beqn\label{tpmf}
&&(P\otimes Q)_1=\left(P_1\otimes Q_0\right)\oplus \left(P_0\otimes Q_1\right)
\overset{r_1}{\underset{r_0}{\rightleftarrows}}
\left(P_0\otimes Q_0\right)\oplus \left(P_1\otimes Q_1\right)=(P\otimes Q)_0\nonumber\\
&&{\rm with}\quad r_1=\left(\begin{array}{cc} p_1\otimes \id_{Q_0} & -\id_{P_0}\otimes q_1 \\ \id_{P_1}\otimes q_0 &
p_0\otimes\id_{Q_1}\end{array}\right)\,,\quad
r_0=\left(\begin{array}{cc} p_0\otimes\id_{Q_0} & \id_{P_1}\otimes q_1 \\ -\id_{P_0}\otimes q_0 &
p_1\otimes\id_{Q_1}\end{array}\right)\,,
\eeqn
regarded as a matrix factorisation over $S\p=\CC[X_i,Z_i]$. 
Here, by abuse of notation the tensor products between $P_i$ and $Q_j$ denote the tensor products $P_i\otimes_{\CC[X_i,Y_i]}\CC[X_i,Y_i,Z_i]\otimes_{\CC[Y_i,Z_i]} Q_j$. Obviously, the matrix factorisation $P\otimes Q$ is a matrix factorisation of $W_1(X_i)-W_2(Y_i)+W_2(Y_i)-W_3(Z_i)=W_1(X_i)-W_3(Z_i)$. The bulk fields $Y_i$ of the theory squeezed in between the original defects give rise to new defect degrees of freedom. Hence, $P*Q$
represents a defect between the LG models with superpotentials $W_1(X_i)$ and $W_3(Z_i)$ which a priori can however be of infinite rank. This happens, because
the tensor product will in general still contain the variables $Y_i$. Interpreting $\BC[X_i, Y_i, Z_i]$ as an infinite dimensional $\BC[X_i,Z_i]$ module gives
infinite rank to matrices that contain $Y_i$. 
As was shown in \cite{Brunner:2007qu}, the factorisations $P*Q$ are 
always equivalent to  finite rank factorisations, provided $P$ and $Q$ are of finite rank.

The generalisation of the matrix factorisation representation of fusion to Landau-Ginzburg orbifolds is straight forward. 
The same arguments as in non-orbifolded Landau-Ginzburg models leads one to
consider the tensor product matrix factorisation $P\otimes Q$. 
But now, $P$ and $Q$ are equivariant with respect to $\Gamma_1\times\Gamma_2$ and $\Gamma_2\times\Gamma_3$ respectively. Thus, $P\otimes Q$ is equivariant with respect to $\Gamma_1\times\Gamma_2\times\Gamma_3$. Again $P\otimes Q$ has to be regarded as ($\Gamma_1\times\Gamma_3$ equivariant) matrix factorisation over
$\CC[X_i,Z_i]$, because the $Y_i$ become new defect degrees of freedom. Just like for the BRST-cohomology, the orbifold 
causes a projection onto $\Gamma_2$-invariant degrees of freedom. Thus
\beq
P*_{\rm orb} Q=(P*Q)^{\Gamma_2}\,.
\eeq
This discussion easily extends to fusion of B-type defects with B-type boundary conditions. For this, one only has to replace the matrix factorisation $Q$ above by a 
$\Gamma_2$-graded matrix factorisation of $W_2(Y_i)$ which represents a boundary condition in the LG model with superpotential $W_2(Y_i)$. $P*_{\rm orb}Q$ is 
then  a $\Gamma_1$-equivariant matrix factorisation of $W_1(X_i)$ and represents a boundary condition in the corresponding LG orbifold.
\subsection{Quantum Symmetry Defects in $X^d/\ZZ_d$}\label{qsdef}
As an example let us discuss defects between one and the same Landau-Ginzburg orbifold with only one chiral superfield $X$, superpotential $W(X)=X^d$ and orbifold group $\Gamma=\ZZ_d$ acting on $X$ by
\beq
X\mapsto \xi^a X\,,\quad a\in\ZZ_d\,,
\eeq
where $\xi$ is an elementary $d$th root of unity.
A simple defect in the unorbifolded LG model is the identity defect which can be represented by the matrix factorisation \cite{Brunner:2007qu}
\beq
P:\quad P_1=S \overset{p_1=(X-Y)}{\underset{p_0=\prod_{i\neq 0}(X-\xi^i Y)}{\rightleftarrows}}S=P_0\,,
\eeq
with $S=\CC[X,Y]$.
To obtain out of this a $\Gamma\times\Gamma$-equivariant matrix factorisation, one can use the orbifold construction described above. The subgroup stabilising $P$ is given by the diagonal subgroup $\Gamma_{\rm diag}\subset\Gamma\times\Gamma$. Thus, the first step is to choose a $\Gamma_{\rm diag}\cong\ZZ_d$ representation on $P$. This is done by specifying the $\ZZ_d$ representation $m$ on the subspace spanned by $1\in P_0\cong S$, which by compatibility with the $S$-action extends to a representation on $P_0$, and which by compatibility with the maps $p_i$ determines a representation on $P_1$. One obtains the
$\Gamma_{\rm diag}$-equivariant matrix factorisation
\beq
S[m+1] \overset{p_1=(X-Y)}{\underset{p_0=\prod_{i\neq 0}(X-\xi^i Y)}{\rightleftarrows}}S[m]\,.
\eeq
The $\Gamma/\Gamma\p\cong\{1\}\times\Gamma$-orbit of this matrix factorisation yields the $\Gamma\times\Gamma$-equivariant matrix factorisation
(we only specify $\wt{p}_1$ here)
\beq
\wt{p}_1=\bigoplus_{i\in\ZZ_d}(X-\xi^i Y):\left(S[m+1]\right)^{\oplus d}\rightarrow\left(S[m]\right)^{\oplus d}\,.
\eeq
The representation of $\Gamma\times\Gamma$ on $\wt{P}_0$ is determined by the representation $\overline{\rho}_0$ 
on the subspace 
$\overline{P}_0=\CC[m]^{\oplus d}\subset(S[m])^{\oplus d}$. It is given by
\beq
(a,b)\in\ZZ_d\times\ZZ_d:\quad \overline{\rho}_0(a,b)=\xi^{am}\id_{\overline{P}_0}+\epsilon^{b-a}\,,
\eeq
where $\epsilon$ is the $d\times d$-matrix defined by $\epsilon_{i,j}=\delta^{(d)}_{i-j-1}$. It is now easy to diagonalise $\overline{\rho}_i$ on $\overline{P}_i$. In the corresponding eigenbasis the equivariant matrix factorisation $\wt{P}^m(X,Y)$ reads
\beq\label{idmf}
\wt{p}_1^m=\left(\begin{array}{cccc} \s X & & & \s -Y \\ \s -Y & \ddots & & \\ &
\s\ddots &\s\ddots & \\&& \s-Y&\s X\end{array}\right):S^d
\left(\begin{array}{c}\s \left[m+1,0\right] \\ \s\vdots
\\\s\left[m+d,-d+1\right]\end{array}\right) 
\longrightarrow S^d
\left(\begin{array}{c}\s\left[m,0\right] \\ \s\vdots
\\\s\left[m+d-1,-d+1\right]\end{array}\right)\,,
\eeq
where now $[\cdot,\cdot]$ denotes an irreducible $\Gamma\times\Gamma$-representation defined on the respective subspaces of $\overline{P}_i$.

What we have seen in particular is that the identity defects $P$ of the unorbifolded
theory breaks up into $d$ different defects $\wt{P}^m$ of the $\ZZ_d$ orbifold.
One expects of course that one of them can be identified as the identity
defect of the orbifold. We will now show that the $\wt{P}^m$ 
realise the $\ZZ_d$ group of quantum symmetries of the orbifold theory.

To see this, we first calculate the fusion of two such defects represented by matrix factorisations $P=\wt{P}^m(X,Y)$ and $Q=\wt{P}^{m\p}(Y,Z)$. As discussed above, the result of the composition is given by $P*_{\rm orb}Q$, the $\ZZ_d$-invariant part of the
tensor product matrix factorisation $P\otimes Q$. Indeed, as in the unorbifolded situation, there is a trick, which simplifies the calculation of the fusion. Namely, the tensor product matrix factorisation is equivalent to the matrix factorisation arising after two steps out of a $\CC[X,Z]$-free two-periodic resolution of the module \cite{Brunner:2007qu}
\beq
M=\coker(p_1\otimes\id_{Q_0},\id_{P_0}\otimes q_1)\,,
\eeq
and the $\ZZ_d$-invariant part is equivalent to the matrix factorisation arising in the same way out of $M^{\ZZ_d}$. In order to calculate $M^{\ZZ_d}$, let us first note that $P_0\otimes Q_0$ is generated over $\widehat{S}=\CC[X,Y,Z]$ by 
$e_{a,b}=e_a^P\otimes e_b^Q$ of $\Gamma^3$-degree $[m+a,-a+m\p+b,-b]$, where $(e_a^P)_{a\in\ZZ_d}$ and 
$(e_b^Q)_{b\in\ZZ_d}$ are generating systems of $P_0$ and $Q_0$ respectively. Generators of $P_0\otimes Q_0$ over $S\p=\CC[X,Z]$ are given by $e_{a,b}^i=Y^ie_{a,b}$. The relations in $M$ coming from $p_1\otimes\id_{Q_0}$ and $\id_{P_0}\otimes q_1$ in this basis read
\beq
e_{a+1,b}^{i+1}=Xe_{a,b}^i\,,\qquad
e_{a,b}^{i+1}=Ze_{a,b+1}^i\,\quad\forall i\geq 0\,.
\eeq
The second of these relations can be used to eliminate all $e_{a,b}^i$ for $i>0$ from the generating system of $M$. The remaining relations then become
\beq
Xe_{a,b}^0=Ze_{a+1,b+1}^0\,.
\eeq
Hence $M$ is generated by $e_{a,b}^0$ subject to these relations. Moreover, $M^{\ZZ_d}$ is generated by those generators, which are $\ZZ_d$-invariant (with respect to the second $\ZZ_d$), \ie $f_a:=e_{a,a-m\p}^0$ subject to the relations
\beq
Xf_a=Zf_{a+1}\,.
\eeq
Moreover, the $\ZZ_d\times\ZZ_d$ degree of $f_a$ is given by $[m+a,m\p-a]$. Therefore 
\beq
M^{\ZZ_d}=\coker(\wt{p}_1^{m+m\p}(X,Z))\,,
\eeq
and $M^{\ZZ_d}$ has a $S\p$-free resolution given by $\wt{P}^{m+m\p}(X,Z)$. Hence for the fusion we obtain
\beq
\wt{P}^m(X,Y)*_{\rm orb}\wt{P}^{m\p}(Y,Z)=\wt{P}^{m+m\p}(X,Z)\,.
\eeq
Indeed the $\wt{P}^m$ form a $\ZZ_d$-group under fusion.

We will now check that the action of the defects on boundary conditions
reproduces the action of the $\ZZ_d$ of quantum symmetries.
Boundary conditions in this model are described by $\ZZ_d$-equivariant matrix factorisations of $W$. 
These can be decomposed into the irreducible ones 
\beq
Q^{(M,N)}(X):\quad Q_1=\CC[X][M+N]\overset{q_1=X^N}{\underset{q_0=X^{d-N}}{\rightleftarrows}} Q_0=\CC[X][M]\,,
\eeq
for $(M,N)\in\II_d=\ZZ_d\times\{0,\ldots,d-1\}$. The quantum symmetries act on these matrix factorisations by shifting 
the $\ZZ_d$-representation label $M$. 
 
To calculate the fusion $P*_{\rm orb} Q$ for $P=\wt{P}^m(X,Y)$ with $Q=Q^{(M,N)}(Y)$ we follow the same path as before.
The module $M=\coker(p_1\otimes\id_{Q_0},\id_{P_0}\otimes q_1)$ is generated over $S\p=\CC[X]$ by $e_a^i=Y^ie_a^P\otimes e^Q$ of $\ZZ_d\times\ZZ_d$-degree $[m+a,-a+M+i]$ with relations
\beq
e_{a+1}^{i+1}=Xe_a^i\,,\qquad 
e_a^{N+i}=0\,,\quad\forall i\geq 0\,.
\eeq
The first set of relations can again be used to reduce the generating system to $e_a^0$, and the remaining relations are
\beq
X^Ne_{a-N}^0=0\,.
\eeq
The only $\ZZ_d$-invariant generator is $e_M^0$, hence 
\beq
M^{\ZZ_d}=\coker(q_1^{(m+M,N)}(X))\,,
\eeq
which therefore has an $S\p$-free resolution given by $Q^{(m+M,N)}(X)$. We arrive at
\beq
\wt{P}^m(X,Y)*_{\rm orb}Q^{(M,N)}(Y)=Q^{(M+m,N)}(X)\,.
\eeq
In particular the defects corresponding to $\wt{P}^m$ are the
generators of the quantum $\ZZ_d$-symmetry in the LG orbifold, and the
one for $m=0$ is the identity defect. 

The construction of these defects (and a more general class of topological defects) is
indeed also straight forward on the level of conformal field theory. We have
presented it in Appendix \ref{cftapp}.
\section{A Special Class of Defects between $X^{d}/\ZZ_d$ and $X^{d\p}/\ZZ_{d\p}$}\label{flowdefects}
In the following we will focus our attention to orbifolds of Landau-Ginzburg models with
one chiral superfield $X$ and superpotential $W=X^d$ for some $d$. The orbifold group
$\Gamma=\ZZ_d$ acts on $X$ by multiplication with $d$th roots of unity. 

In this section
we will define a special class of B-type supersymmetric 
defects between such models by constructing a class of 
$\ZZ_{d\p}\times\ZZ_{d}$-graded matrix
factorisations of $Y^{d\p}-X^d$. 
We will then determine their fusion among themselves and 
their fusion with B-type boundary conditions.
\subsection{Construction}
The matrix factorisations defining the special defects are determined
by irreducible representations $m$ of $\ZZ_{d}$ and a $d\p$-tuple of integers $n=(n_0,\ldots,n_{d\p-1})$, $n_i\in\NN_0$ such that $\sum_in_i=d$. 
We will denote the set of all such pairs $(m,n)$ by $\II_{d\p,d}$. Given an $n$ as above, define the following $d\p\times d\p$-matrix 
\beq
(\Xi_{n})_{a,b}:=\delta^{(d\p)}_{a,b+1}X^{n_a}\,.
\eeq
This matrix has the property that 
\beq
\Xi_n^{d\p}=X^d\id_{d\p}
\eeq
and hence can be used to construct matrix factorisations of $Y^{d\p}-X^d$ by means of 
\beq
(Y^{d\p}-X^d)\id_{d\p}=\prod_{i=0}^{d\p-1}(Y\id_{d\p}-\xi^i\Xi_n)\,,
\eeq
where $\xi$ is an elementary $d\p$th root of unity. In particular choosing a subset $I\subset\{0,\ldots,d\p-1\}$ one obtains a 
matrix factorisation of $Y^{d\p}-X^d$ by grouping together the corresponding factors into one matrix and the ones corresponding to
the complement into the other:
\beq\label{ungradedmf}
p_1=\prod_{i\in I}(Y\id_{d\p}-\xi^i\Xi_n)\,,\quad
p_0=\prod_{i\in\{0,\ldots,d\p-1\}-I}(Y\id_{d\p}-\xi^i\Xi_n)\,.
\eeq
This matrix factorisation is $\ZZ_{d\p}\times\ZZ_{d}$ gradable. The grading is determined by the grading of a single factor 
$(Y\id_{d\p}-\xi^i\Xi_n)$. In particular, the grading of any matrix factorisation of type \eq{ungradedmf} can be obtained from the grading of the matrix factorisation with $I=\{0\}$ on which we will focus now.
For $I=\{0\}$, given $(m,n)\in\II_{d\p,d}$, the respective graded matrix factorisation
$P^{(m,n)}=P^{(m,n)}_{\{0\}}$ is defined by
\beq\label{mfdef}
p_1^{(m,n)}=(Y\id_{d\p}-\Xi_n)=
\left(\begin{array}{cccc} \s Y & & & \s -X^{n_0} \\ \s -X^{n_1} & \ddots & & \\ &
\s\ddots &\s\ddots & \\&& \s -X^{n_{d\p-1}}&\s  Y\end{array}\right):P_1\longrightarrow P_0\,,
\eeq
where
\beq
P_1=S^{d\p}\left(\begin{array}{c}\s\left[1,-m\right] \\ \s\left[2,-m-n_1\right] \\ \s\left[3,-m-n_1-n_2\right]\\
\s\vdots
\\\s\left[d\p,-m-\sum_{i=1}^{d\p-1}n_i\right]\end{array}\right)\,,\qquad
P_0=S^{d\p}\left(\begin{array}{c}\s\left[0,-m\right] \\ \s\left[1,-m-n_1\right] \\ \s\left[2,-m-n_1-n_2\right]\\
\s\vdots
\\\s\left[d\p-1,-m-\sum_{i=1}^{d\p-1}n_i\right]\end{array}\right)\,.
\eeq
Here $S=\CC[X,Y]$, and $[\cdot,\cdot]$ denotes the $\ZZ_{d\p}\times\ZZ_{d}$-degree. 
Note that because we are in the orbifold category, symmetry operations $X\mapsto \eta^i X$, $Y\mapsto \xi^j Y$, where 
$\eta$ is an elementary $d$th root of unity act trivially on the matrix factorisations above. 
\subsection{Fusion of Defects}
Let us consider two matrix factorisations  of the type defined in \eq{mfdef}.
For $(m,n)\in\II_{d\p,d}$ and $(\wt{m},\wt{n})\in\II_{d\p\p,d\p}$ let 
\beq
P:=P^{(m,n)}(Y,X)\,,\qquad
Q:=P^{(\wt{m},\wt{n})}(Z,Y)
\eeq
be the respective graded matrix factorisations of $Y^{d\p}-X^d$ and $Z^{d\p\p}-Y^{d\p}$ respectively.
We would like to calculate the fusion of the respective defects. As discussed in Section \ref{deflgorb} the fused defect can be represented by the $\ZZ_{d\p}$-invariant part of the tensor product matrix factorisation $P\otimes Q$ regarded as matrix factorisation over $S\p:=\CC[X,Z]$. By the usual trick \cite{Brunner:2007qu} which has already been used in the discussion of the fusion of the quantum symmetry defects in Section \ref{qsdef} it can be obtained as the matrix factorisation
associated to the $\ZZ_{d\p}$-invariant part of the 
module 
\beq
M=\coker(p_1\otimes\id_{Q_0},\id_{P_0}\otimes q_1)\,.
\eeq
Here as in Section \ref{qsdef} above, by abuse of notation $P_i\otimes Q_j$ denotes the tensor product
over $\widehat{S}=\CC[X,Y,Z]$ of the respective $\widehat{S}$-modules
$P_i\otimes_{\CC[X,Y]}\widehat{S}$ and $Q_i\otimes_{\CC[Y,Z]}\widehat{S}$, and 
$M$ is regarded as an $S\p$-module.

In order to analyse $M$, let us denote by $(e^P_a)_{a\in\ZZ_{d\p}}$ the free generators of $P_0$ of $\ZZ_{d\p}\times\ZZ_{d}$-degrees $[e^P_a]=[a,-m-\sum_{i=1}^a n_i]$, and by $(e^Q_b)_{b\in\ZZ_{d\p\p}}$ the generators of $Q_0$ with $\ZZ_{d\p\p}\times\ZZ_{d\p}$-degree $[e^Q_b]=[b,-\wt{m}-\sum_{i=1}^b \wt{n}_i]$. We define the corresponding generators
$e_{a,b}:=e^P_a\otimes e^Q_b$ of $P_0\otimes Q_0$ of $\ZZ_{d\p\p}\times\ZZ_{d\p}\times\ZZ_{d}$-degree
\beq\label{gendegree}
[e_{a,b}]=[b,a-\wt{m}-\sum_{i=1}^b \wt{n}_i,-m-\sum_{i=1}^a n_i]\,.
\eeq
As an $S\p=\CC[X,Z]$-module, $P_0\otimes Q_0$ is generated by $e_{a,b}^j:=Y^je_{a,b}$. In this basis the relations
in $M$ coming from $p_1\otimes\id_{Q_0}$ can be written as
\beq\label{relp}
e_{a,b}^{j+1}=X^{n_{a+1}}e_{a+1,b}^j\quad\forall j\in\NN_0\,.
\eeq
They imply
\beq\label{elimination}
e_{a,b}^j=X^{\sum_{i=1}^jn_{a+i}}e_{a+j,b}^0\,,
\eeq
and can hence be used to eliminate $e_{a,b}^j$ with $j>0$ from the generating system of $M$.

The relations coming from $\id_{P_0}\otimes q_1$ on the other hand read
\beq
Ze_{a,b}^j=e_{a,b+1}^{j+\wt{n}_{b+1}}\quad\forall j\in\NN_0\,.
\eeq
Using \eq{elimination} they become
\beq\label{relqel}
ZX^{\sum_{i=1}^jn_{a+i}}e_{a+j,b}^0=X^{\sum_{i=1}^{j+\wt{n}_{b+1}}n_{a+i}}e_{a+j+\wt{n}_{b+1},b+1}^0\quad\forall j\in\NN_0\,.
\eeq
Obviously, the relations \eq{relqel} for $j>0$ follow from the ones with $j=0$, so that $M$ is isomorphic to
the $S\p$-module generated by $e_{a,b}^0$ subject to the relations 
\beq\label{fusrel}
Ze_{a,b}^0=X^{\sum_{i=1}^{\wt{n}_{b+1}}n_{a+i}}e_{a+\wt{n}_{b+1},b+1}^0\,.
\eeq
In particular, $M^{\ZZ_{d\p}}$ is isomorphic to the $S\p$-module generated by 
the $\ZZ_{d\p}$-invariant generators
\beq
f_{b}:=e_{\wt{m}+\sum_{j=1}^b\wt{n}_j,b}
\eeq
subject to the relations
\beq
Zf_b=X^{\sum_{i=1}^{\wt{n}_{b+1}}n_{\wt{m}+\sum_{j=1}^b \wt{n}_j+i}}f_{b+1}\,.
\eeq
These relations can indeed be represented by a matrix of type \eq{mfdef}. More precisely
\beq
M^{\ZZ_{d\p}}\cong\coker\left(p_1^{(\widehat{m},\widehat{n})}(X,Z)\right)
\eeq
with $(\widehat{m},\widehat{n})\in\II_{d\p\p,d}$ given by
\beq\label{labelfusion}
\widehat{m}=m+\sum_{i=1}^{\wt{m}}n_i\,,\quad
\widehat{n}_{b+1}=\sum_{i=1}^{\wt{n}_{b+1}} n_{\wt{m}+\sum_{j=1}^{b}\wt{n}_j+i}\,.
\eeq
Therefore, the class of defects defined by matrix factorisations \eq{mfdef} is closed under fusion. For every $(m,n)\in\II_{d\p,d}$ 
and $(\wt{m},\wt{n})\in\II_{d\p\p,d\p}$
fusion is given by
\beq\label{pfusion}
P^{(\wt{m},\wt{n})}*P^{(m,n)}=P^{(\widehat{m},\widehat{n})}\,,
\eeq
where $(\widehat{m},\widehat{n})=:(\wt{m},\wt{n})*(m,n)\in\II_{d\p\p,d}$ 
is defined by \eq{labelfusion}. Indeed, it is not difficult to see that for every $(\widehat{m},\widehat{n})\in\II_{d\p,d}$
there exist $(m_i,n_i)\in\II_{d+i+1,d+i}$, $0\leq i<d\p-d$ such that 
\beq
(\widehat{m},\widehat{n})=(m_{d\p-d},n_{d\p-d})*\ldots*(m_0,n_0).
\eeq
That means that every defect $P^{(\widehat{m},\widehat{n})}$ between Landau-Ginzburg orbifolds $X^{d}/\ZZ_d$ and $X^{d\p}/\ZZ_{d\p}$ can be obtained by fusion of $|d\p-d|$ defects between Landau-Ginzburg orbifolds with $|d-d\p|=1$. 
This will become more evident using a pictorial representation of these defects which will be introduced in Section \ref{picture} below after the discussion of their action on boundary conditions.

It is also easy to calculate the fusion of the defects $P^{(m,n)}$ with the defects $\wt{P}^m$ representing the quantum symmetries. 
One obtains
\beqn
&&\wt{P}^{m\p\p}*P^{(m,n)}*\wt{P}^{m\p}=P^{(\widehat{m},\widehat{n})}\,,\\
&&\widehat{m}=m+m\p+\sum_{j=1}^{\{-m\p\p\}_{d\p}}n_j\,,\quad
\widehat{n}=(\widehat{n}_0,\ldots,\widehat{n}_{d\p})=(n_{-m\p\p},n_{-m\p\p+1},\ldots, n_{d\p-m\p\p-1})\,.\nonumber
\eeqn
\subsection{Fusion of Defects and Boundary Conditions}
Next, we would like to calculate what happens to B-type boundary conditions in Landau-Ginzburg orbifolds
$X^{d}/\ZZ_{d}$ upon fusion with a defect represented by $P^{(m,n)}$ for some $(m,n)\in\II_{d\p,d}$.
B-type boundary conditions in this model can be represented by $\ZZ_d$-graded matrix factorisations of $X^d$. As already 
mentioned in Section \ref{deflgorb}, 
the latter
can be decomposed into sums of the irreducible matrix factorisations
\beq
Q^{(M,N)}:\quad Q_1=\CC[X][M+N]\overset{q_1=X^N}{\underset{q_0=X^{d-N}}{\rightleftarrows}} Q_0=\CC[X][M]\,,
\eeq
for $(M,N)\in\II_d=\ZZ_d\times\{0,\ldots,d-1\}$. 
Thus, it is sufficient to determine the fusion of $P^{(m,n)}$ with these. 

Similar to the case of fusion of defects also the boundary condition created by fusing the defect associated to 
$P=P^{(m,n)}$ with the boundary condition associated to $Q=Q^{(M,N)}$ is represented by the matrix factorisation
obtained from the $\ZZ_d$-invariant submodule of 
\beq
M=\coker(p_1\otimes\id_{Q_0},\id_{P_0}\otimes q_1)
\eeq
regarded as $S\p=\CC[Y]$-module. 
We denote the $S=\CC[X,Y]$-free generators of $P_0\otimes Q_0$ by 
$e_a$, ${a\in\ZZ_{d\p}}$. They have  $\ZZ_{d\p}\times\ZZ_{d}$-degree 
$[a,-m-\sum_{i=1}^a n_i+M]$. $S\p$-free generators of 
$P_0\otimes Q_0$ are given by $e_a^i=X^ie_a$, $i\geq 0$.
In this generators, the relations in $M$ can be written as
\beq
Ye_a^i=e_{a+1}^{i+n_{a+1}}\,,\quad
e_a^{N+i}=0\,.
\eeq
By means of these relations, one can reduce the set of generators to those
$e_a^i$ with $0\leq i\leq \min(N,n_a)-1$. The $\ZZ_d$-invariant ones are the ones with
\beq\label{imap}
i=i(a)=\{m-M+\sum_{j=1}^a n_j\}_d\,,
\eeq
where $\{z\}_d$ denotes the representative in $\ZZ$ of $z\in\ZZ_d$ which lies in $[0,d-1]$.
A generator  $e_a^{i(a)}$ contributes to $M^{\ZZ_d}$ iff $i(a)<\min(N,n_a)$.
Using the relation 
\beq
Y^ke_a^i=e_{a+k}^{i+\sum_{j=1}^kn_{a+j}}
\eeq
one easily obtains that $e_a^{i(a)}$ generates a submodule with relation
\beq
Y^k e_a^{i(a)}=0\,,\quad \forall k:\, i(a)+\sum_{j=1}^kn_{a+j}\geq N\,.
\eeq
The $\ZZ_{d\p}$-degree of this generator is given by $[e_a^{i(a)}]=[a]$. Hence
\beq
M^{\ZZ_d}\cong\bigoplus_{{a\in\ZZ_{d\p}}\,:\,{i(a)=\{m-M+\sum_{j=1}^a n_j\}_d<\min(N,n_a)}}\coker\left(q_1^{(a,k(a))}\right)\,,
\eeq
where 
\beq\label{kdef}
k(a)=\min\{j>0\,|\,i(a)+\sum_{k=1}^jn_{a+k}\geq N\}\,.
\eeq
In particular, the fusion reads
\beq\label{Paction}
P^{(m,n)}*Q^{(M,N)}=\bigoplus_{{a\in\ZZ_{d\p}}\,:\,{i(a)=\{m-M+\sum_{j=1}^a n_j\}_d<\min(N,n_a)}}
Q^{(a,k(a))}\,.
\eeq
Indeed, for all $(m,n)\in\II_{d\p,d}$ and $(M,N)\in\II_d$ this sum has at most one summand.
This can easily be seen as follows. Suppose $i(a)<n_a$ for some $a\in\ZZ_{d\p}$, giving rise to a possible summand in \eq{Paction}. Then 
\beq
i(a\p)=i(a)+\sum_{j=1}^{\{a\p-a\}_{d\p}}n_{a+j}\,,
\eeq
because from $i(a)<n_a$ it follows that the right hand side is $<d$ for all $a\p$. Since now all the summands are non-negative this implies that $i(a\p)\geq n_{a\p}$ for all $a\p\neq a$, and therefore no $a\p\neq a$ can contribute to the sum in \eq{Paction}.

However, the sum in \eq{Paction} can be empty if there exist $n_i\geq 2$. More precisely, for each
$n_i\geq 2$ matrix factorisations $Q=Q^{(M,N)}$ are annihilated\footnote{Note that since the supersymmetric boundary conditions in the models at hand are classified, one immediately obtains that the fusion of the respective defects and boundary conditions also vanishes identically in the full conformal field theory, provided this fusion is regularised in a  supersymmetric way.} 
by $P^{(m,n)}$, iff 
 \beq
 N\leq n_i-1\,,M\in(m+1+n_1+\ldots+n_{i-1})+\{0,\ldots,n_i-N-1\}\,.
 \eeq
 This can be seen by considering the set
  $\JJ:=\{i(a)\,|\,a\in\ZZ_{d\p}\}$ of possible values of $i(a)$. $\JJ$ is a subset of $\ZZ_d$, and its complement
  is given by 
  \beq
  \JJ^c=(m-M+1)+\Big([0,n_1-2]\cup (n_1+[0,n_2-2])\cup\ldots\cup(n_1+\ldots+n_{d-1}+[0,n_d-2])\Big)\,.
  \eeq
In particular for $M=m+1+n_1+\ldots+n_{i-1}+r$, $0\leq r\leq n_i-N-1$ 
\beqn
\JJ^c&=&(-n_1-\ldots-n_{i-1}-r+[0,n_1-2])\cup\ldots\cup (-r+[0,n_i-2])\cup\\
&&\qquad\qquad\qquad\qquad\qquad\qquad\cup\ldots\cup (-r+n_i+\ldots+n_{d-1}+[0,n_{d}-2])\,.\nonumber
\eeqn
But this means that for all $a\in\ZZ_{d\p}$ $i(a)> n_i-2-r\geq N-1$, \ie $i(a)\geq N$ for all $a$, and hence the sum in
\eq{Paction} is empty.

Let us suppose now, that $Q^{(M,N)}$ is not annihilated. This means that 
\beq\label{nacond}
\JJ\cap\{0,\ldots,N-1\}=\{i_1,\ldots,i_l\}\neq\emptyset\,.
\eeq
Assume $i_1=i(a_1)$ is the smallest of the $i_j$ (considered as elements of $\ZZ$ in the range $\{0,\ldots,d-1\}$). Then, $i_1<n_{a_1}$, because otherwise $0\leq i_1-n_{a_1}<i_1$ would also be an element of the set above. Hence
\beq\label{Paction2}
P^{(m,n)}*Q^{(M,N)}=Q^{(a_1,k(a_1))}\,.
\eeq
This formula looks rather implicit, but there is a nice pictorial way to understand it, which we will discuss in the next Section.
\subsection{Pictorial Representation of Defect Action}\label{picture}
Let us for the moment restrict the discussion to those $P=P^{(m,n)}$ with $n_i\geq 1$ for all $i$. This implies in particular
$d\geq d\p$. Obviously this property is preserved under fusion. 
The first thing to note is that under this assumption the action of $P$ on a matrix factorisation $Q^{(M,N)}$ does not increase $N$, which is obvious from \eq{kdef}.
This implies in particular that under the fusion with $P$, $Q^{(M,1)}$ is either annihilated or it is 
mapped to $Q^{(M\p,1)}$ for some $M\p\in\ZZ_{d\p}$. 
The ones which are not annihilated are the ones such that there exists an $a\in\ZZ_{d\p}$ with $i(a)=0$, \ie
those with 
\beq
M\in m+\{0,n_1,n_1+n_2,\ldots,n_1+\ldots+n_{d\p-1}\}=:{\cal L}_{(m,n)}\,,
\eeq
and for $M=m+\sum_{j=1}^an_j$ one obtains $M\p=a$. 
Summarising, for the action of $P$ on the $N=1$ boundary conditions we get
\beq\label{N=1sum}
P^{(m,n)}*Q^{(M,1)}=\left\{\begin{array}{ll} 0\,,& {\rm if}\;M\notin m+\{\sum_{i=1}^a n_i\,|\,0\leq a<d\p\} \\
Q^{(a,1)}\,, & {\rm if}\; M=m+\sum_{i=1}^a n_i\end{array}\right.\,.
\eeq
This suggest the following picture for the action of the defects $P$ on the $Q^{(M,1)}$.
\FIGURE[ht]{
\includegraphics[width=100mm]{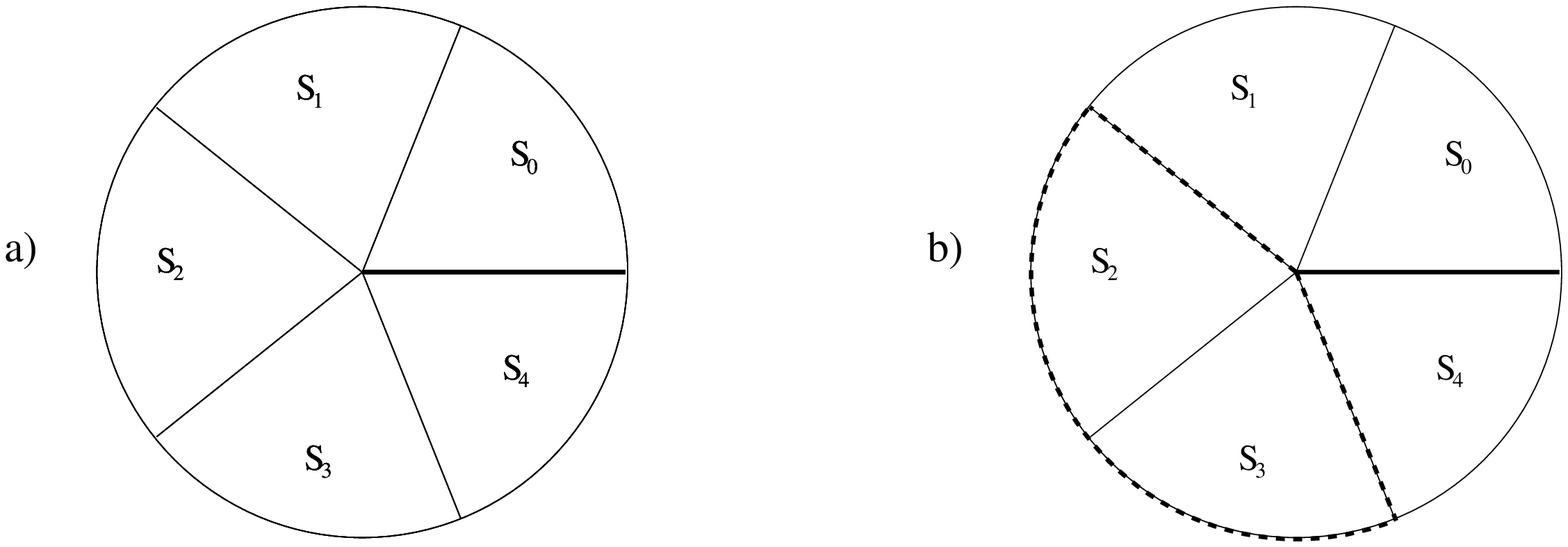}
\caption{a) Disk subdivided into $d$ sectors $S_i$ representing boundary conditions $Q^{(i,1)}$.
b) Union of consecutive sectors represent boundary conditions $Q^{(M,N)}$ with $N>1$, \eg $S_2\cup S_3$ representing boundary condition $Q^{(2,2)}$.\label{fig1}}
}
Consider a disk
subdivided by straight lines from its center to its boundary into $d$ sectors. Mark one of the straight lines, and denote the sectors by $S_0$ to $S_{d-1}$ going in counterclockwise direction and starting from the marked line (\cf Figure \ref{fig1}a).
In this picture we represent matrix factorisations $Q^{(M,1)}$ by the $M$th sector $S_M$. 

Now we can consider the following pictorial operations. The first rather trivial one ${\cal T}_m$ is the shift of the marking to the $-m$th line in counterclockwise direction, which just corresponds to the quantum symmetry $Q^{(M,1)}\mapsto Q^{(M+m,1)}$ 
(\cf Figure \ref{fig2}a). A more interesting operation is the operation ${\cal S}_{\{s_1,\ldots,s_{d-d\p}\}}$,
which shrinks to zero the sectors $S_{s_i}$
by bringing together the lines bounding them. In this way, from a disk subdivided into $d$ sectors $S_{M}$ one obtains
a disk 
subdivided into $d\p$ sectors $S\p_{M\p}$ again counted in counterclockwise direction from the marked line (\cf Figure \ref{fig2}b). By means of the identification 
of boundary conditions $Q^{(M,1)}$ with sectors $S_M$ the operation of $P^{(m,n)}$ in \eq{N=1sum} can be written as
\beq\label{pictaction}
{\cal O}^{(m,n)}={\cal S}_{{\cal L}^c_{(m,n)}-m}{\cal T}_{-m}={\cal T}_{-a_{(m,n)}}{\cal S}_{{\cal L}^c_{(m,n)}}\,,
\eeq
where ${\cal L}^c_{(m,n)}$ is the complement of the set ${\cal L}_{(m,n)}$ 
of $\ZZ_d$-labels of the 
non-annihilated $N=1$-boundary conditions. $a_{(m,n)}:=|\{0,\ldots, m\}\cap {\cal L}_{(m,n)}|$ is the number of
segments before the $m$th one which are not shrunken. 
\FIGURE[ht]{
\includegraphics[width=100mm]{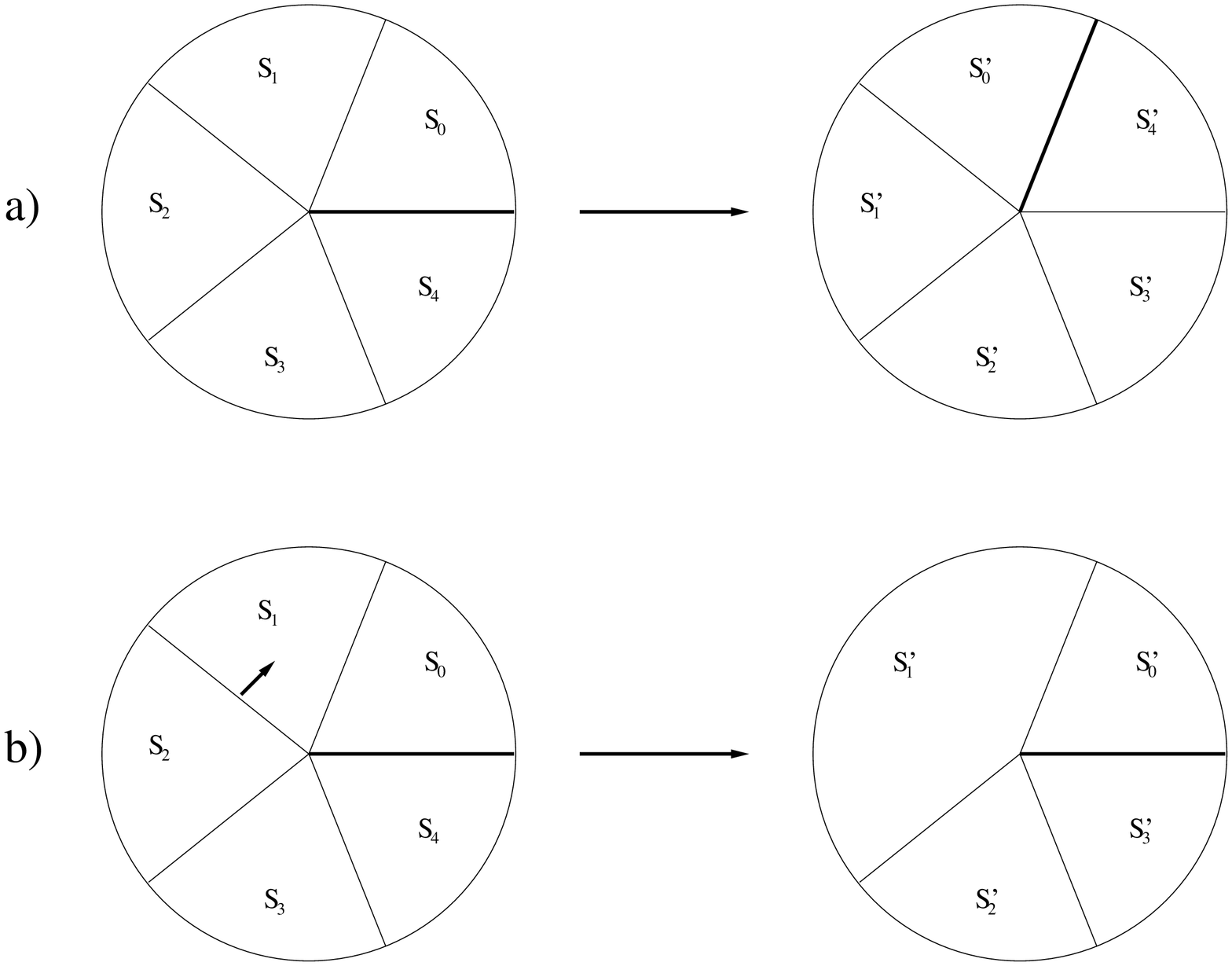}
\caption{Diskoperations: a) ${\cal T}_{-1}$: marked line shifted by $1$, $S_i\mapsto S\p_{i-1}$,
b) ${\cal S}_{\{1\}}$: sector $S_1$ shrunken to zero, $S_0\mapsto S\p_0$, $S_1\mapsto 0$, $S_i\mapsto S\p_{i-1}$ for $1<i\leq 4$.}\label{fig2}}

Indeed, this pictorial representation of the action of $P^{(m,n)}$ generalises to the action on all boundary conditions 
if one represents $Q^{(M,N)}$ for arbitrary $N$ by 
the union\footnote{The decomposition of such a union into its constituents indeed corresponds to the D-brane charge decomposition.} of the sectors $S_M\cup S_{M+1}\cup \ldots\cup S_{M+N-1}$ (\cf Figure \ref{fig1}b). This can be seen as follows.
Consider first the situation, in which the pictorial operation deletes all the sectors belonging to the pictorial representation of a 
given boundary condition $Q^{(M,N)}$, \ie
the set $\{M,M+1,\ldots, M+N-1\}$ is completely contained in ${\cal L}^c_{(m,n)}$. 
In this case, by definition, $i(a)\geq N$ for all $a$, and hence by \eq{Paction} $Q^{(M,N)}$ is annihilated by $P$. Thus, the
pictorial action  \eq{pictaction} agrees with the action of $P$. If on the other hand, the pictorial operation does not
delete all segments belonging to the boundary condition $Q^{(M,N)}$, then as in  \eq{nacond}
\beq
{\cal L}_{(m,n)}\cap\{M,M+1,\ldots,M+N-1\}=\JJ\cap\{0,\ldots,N-1\}=\{i_1,\ldots,i_l\}\neq\emptyset\,,
\eeq
and $P$ does not annihilate $Q^{(M,N)}$. The result of the fusion has already been stated in \eq{Paction2}. Since $n_i\geq 1$ for all $i$, we obviously obtain $k(a_1)=l$. But this is exactly the number of those segments of the pictorial representation of $Q^{(M,N)}$, which are not annihilated by $P$. 
Furthermore, $a_1$ is the number of $Q^{(M\p,1)}$
with $M\p\in\{m,\ldots,M\}$ which are not annihilated by $P$. Thus, also in this case \eq{pictaction} applied to the pictorial representation of $Q^{(M,N)}$ is nothing but the pictorial representation of the result \eq{Paction2} of the fusion of $P$ and $Q^{(M.N)}$. This shows that indeed ${\cal O}^{(m,n)}$ represents the action of $P^{(m,n)}$ on all boundary conditions.

In fact, a similar picture also describes the action of 
$P=P^{(m,n)}$ where $n_a=0$ is allowed. In this case one has to replace the pictorial action \eq{pictaction} by
\beq\label{pictactiongen}
\wt{\cal O}^{(m,n)}=\wt{\cal S}_{(m,n)}{\cal T}_{-m}\,,
\eeq
where now $\wt{\cal S}_{(m,n)}$ not only deletes all the segments $S_M$ for which $M$ is not in the image of the map
\beq
\wt{\imath}(a)=\sum_{j=1}^an_j\,,
\eeq
but in addition it also splits up every segment $S_M$ into $|\wt{\imath}^{-1}(M)|$ segments. Thus, not only are $n_i-1$ segments 
deleted for each $i$ with $n_i>1$, but also a new segment is created for each $i$ with $n_i=0$. For the flows
between minimal model orbifolds however, only those defects $P^{(m,n)}$ with $n_i\geq 1$ play a role. 
\section{Defects and Bulk Flows between Minimal Model Orbifolds}\label{comparison}
We propose that the defects presented in Section \ref{flowdefects} above 
arise in the way described in Section \ref{bulkflows} in supersymmetric bulk flows between orbifolds 
${\cal M}_{d-2}/\ZZ_d$ of $N=2$ superconformal minimal models.  
To give evidence for this proposal, we will analyse these flows in the mirror Landau-Ginzburg models in the following, and compare them
to the fusion of the defects $P^{(m,n)}$ calculated in Section \ref{flowdefects}.
\subsection{Flows in the Mirror Landau-Ginzburg Models}
As mentioned in Section \ref{setsum}, in the mirror LG models, the flows we are interested in correspond to
lower order deformations $W_\lambda$ of the superpotential $W=W_{\lambda=0}=X^d$. We would like to 
describe what happens to the corresponding A-type D-branes under such deformations. The relevant information
about this is encoded in the structure of the critical points of the superpotential. 

Let $p$ be any polynomial of degree $d$ in one variable. Regarded as a map $\CC\rightarrow\CC$, it is a $d$-sheeted branched cover of 
the complex plane. The branch points are the critical points $x_i^*$ of $p$, in which $o(x_i^*)+1$ many sheets meet. Here $o(x_i^*)$ denotes the order of the critical point.
Let us choose a base point $b$ near $\infty$ in the image of $p$, which is not a critical value. The preimage $p^{-1}(b)$ consists of $d$ points which we denote by $b_a$, $a\in\ZZ_d$ in such a way that the monodromy around $\infty$ acts on the fiber over $b$ by $b_a\mapsto b_{a+1}$. 
  
 \FIGURE[ht]{
\includegraphics[width=150mm]{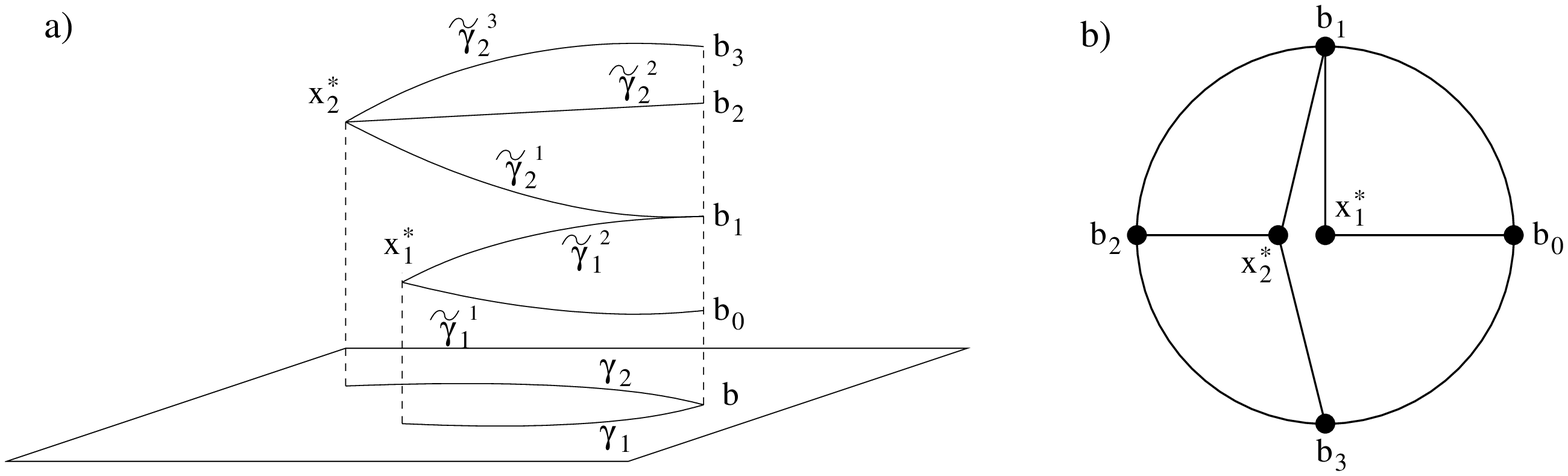}
\caption{Polynomial of degree $4$ with two critical points $x_1^*$ and $x_2^*$ of degrees $o(x_1^*)=1$ and $o(x_2^*)=2$
respectively. a) Paths $\gamma_i$ between critical values $p(x_i^*)$ and base point $b$ and their lifts $\wt{\gamma}_i^\mu$ to the preimage of $p$. b) Schematic representation.}\label{pathsfig}
}
Now let us suppose that all the critical values of $p$ are different, and choose paths $\gamma_i$ 
from the critical values $p(x_i^*)$ to $b$ which only intersect each other in $b$. Then the preimage $p^{-1}(\gamma_i)$
consists of $o(x_i^*)+1$ paths $\wt{\gamma}_i^\mu$ going from $x_i^*$ to $o(x_i^*)+1$ distinct preimages $b_{a_i^\mu}$ of $b$ (\cf Figure \ref{pathsfig}a). 
Taking $b$ to $\infty$ and compactifying $\CC$ to the disk, 
we obtain the following schematic representation (\cf Figure \ref{pathsfig}b).
The points $b_a$ are distinct points on the boundary of the disk, which are cyclically ordered, and each of the critical points
$x_i$ in the interior of the disk is connected by the $\wt{\gamma}_i^\mu$ to $o(x_i^*)+1$ of them. We call the union of these paths $\Gamma_i$. $\Gamma_i$ and $\Gamma_j$ for $i\neq j$ can only intersect on the boundary of the disk. 
Since $\sum_io(x_i^*)=d-1$
and all $b_a$ have to be connected to each other on $\Gamma=\bigcup\Gamma_i$, 
each $b_a$ can only lie on at most two different $\Gamma_i$, and $\Gamma$ has to be simply connected, \ie
there are no closed loops on it.

Note however that this graphical representation
depends on a choice of the (homotopy class of the) paths $\gamma_i$. In the following we will make a choice which is adapted to 
the description of A-branes in Landau-Ginzburg models. The latter are one-dimensional submanifolds of $\CC$ on which the imaginary part $\Im(W)$
is constant and on which the real part $\Re(W)$ is bounded from below \cite{Hori:2000ck}. This means in particular that the world volumes of A-branes are unions $(-\wt{\gamma}_i^\mu)\cup \wt{\gamma}_i^{\mu\p}$ of preimages under $W$ 
of paths $\gamma_i=W(x_i^*)+\RR^{\geq 0}$, where now $x_i^*$ are the critical points of $W$. (A minus sign in front of
a path indicates the inversion of the parametrisation or orientation.)
Thus, if we assume that $\Im(W(x_i^*))\neq\Im(W(x_j^*))$ for all $i\neq j$, this choice of paths $\gamma_i$ gives rise to a schematic representation of A-branes
in the LG model.

For instance for $W=X^d$, there is one critical point $x^*=0$ of order $d-1$. The critical value $W(x^*)=0$, thus A-branes consist 
of unions of two different 
premiages under $W$ of the nonnegative real line $\RR^{\geq 0}$, 
which are just $\wt{\gamma}^\mu=e^{2\pi i\mu\over d}\RR^{\geq 0}$ for $\mu\in\{0,\ldots, d-1\}$.  The graphical representation is hence a disk with one point
in the interior from which $d$ lines representing $\wt{\gamma}^\mu$
go to the points $b_\mu$ on the boundary, and A-branes are unions
$(-\wt{\gamma}^\mu)\cup\wt{\gamma}^\nu$ which we will denote by $\overline{b_\mu x^* b_\nu}$.

Under a deformation $W_\lambda$ of $W$, the critical point $x^*$ splits up into $N$ distinct critical points $x_i^*$. If we assume that for all $\lambda>0$ the imaginary parts $\Im(W(x_i^*))$ are all distinct, and no further splitting of critical points occurs, then the "topology" of the graphical representation does not change. 

The renormalisation group flow  now drives $W_\lambda$ to a homogeneous superpotential, \ie at its endpoint, there is only a single critical point left at $0$. The other critical points 
go off to $\infty$. If under the RG flow the imaginary parts $\Im(W(x_i^*))$ of the critical values all stay separate
and no further splitting of critical points occur, than it is easy to see what happens to A-branes under this perturbation. 
A-branes which are attached to critical points $x_i^*$, $i>1$ going off to $\infty$ 
decouple\footnote{Their bulk-boundary couplings go to zero in the IR. This is clear for the topological couplings. Since the flows at hand preserve supersymmetry also in the boundary sectors, and since furthermore  
the supersymmetric boundary conditions in these models are classified, it also follows for all bulk-boundary couplings on the level of the full conformal field theory.}
 from the theory, while A-branes attached to the critical point $x_1^*$ which remains finite flow to the respective A-branes in the IR.
A-branes consisting of rays which are separated by the perturbation, \ie rays which emanate from different critical points for $\lambda\neq 0$ have to decay into sums of A-branes of the two types above by addition and subtraction\footnote{Addition with opposite orientation.} of rays going to those boundary points $b_a$ which lie on intersections of graphs $\Gamma_i$ and $\Gamma_j$. The summands then behave as described above. For the special class of perturbations $W_\lambda=X^{nd}+\lambda X^d$ this has been analysed in detail in \cite{Gaberdiel:2007us}.

This flow on A-branes has a simple description in terms of the graphical representation of the deformations $W_\lambda$. In the UV, A-branes $\overline{b_ix^*b_j}$ are specified by pairs $(b_{i},b_{j})$ of two different boundary points.
The same is true in the IR, where however only the boundary points $b_{a_1^\mu}$ remain. The flow associated to a graphical representation on the level of A-branes is then just described by identifying all boundary points $b_i\sim b_j$ which are connected on $\Gamma-\Gamma_1$. An A-brane $(b_i,b_j)$
in the UV therefore flows to the brane $([b_i],[b_j])$ in the IR, where $[\cdot]$ denotes the equivalence class 
with respect to the equivalence relation $\sim$. If in particular 
the two points $(b_i,b_j)$ defining an A-brane in the UV are identified by $\sim$ then the brane decouples from the theory.  Note that while the set $\{[b_i]\}$ of rest classes forms 
a cyclically ordered set, there is an ambiguity of identifying it with $\ZZ_{d\p}$. The latter is related to the freedom of a quantum symmetry operation in the IR.
\FIGURE[ht]{
\includegraphics[width=100mm]{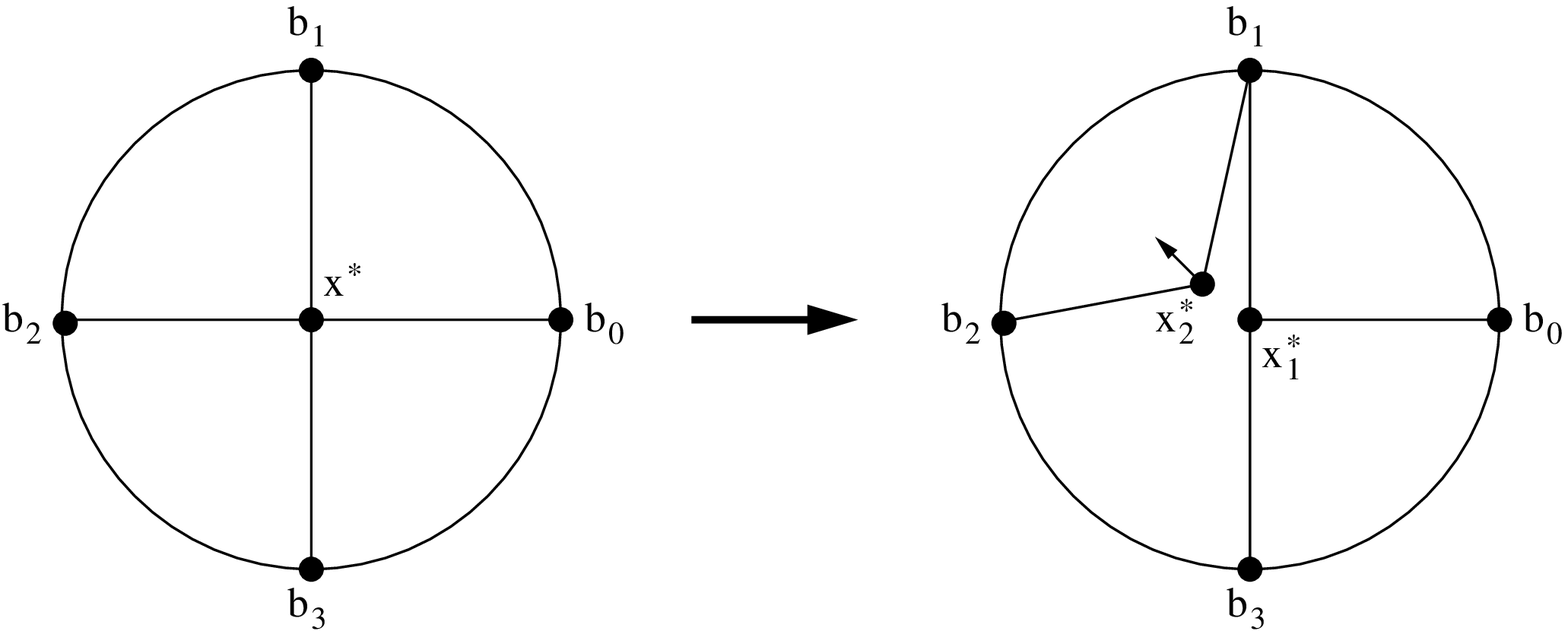}
\caption{Flow corresponding to $W_\lambda=X^4+\lambda X^3$ for a particular choice of $\lambda$.}
\label{pertfig}
}

As a simple example let us consider the perturbation corresponding to $W_\lambda=X^d+{\lambda}X^{d-1}$. The corresponding RG flow drives the system from the LG model with superpotential $W=X^d$ to the one with $W=X^{d-1}$ (\cf Section \ref{setsum}). The critical points of $W_\lambda$ are
$x_1^*=0$ of order $d-2$ and $x_2^*=-\lambda$ of order $1$. Thus, for $\lambda\neq 0$ 
a critical point $x_2^*$ of order $1$ splits off from the
critical point in $0$ and goes to $\infty$ under the RG flow. The A-brane $\overline{b_{a_2^1}x_2^*b_{a_2^2}}$ consisting
of the two preimages $\wt{\gamma}_2^1$ and $\wt{\gamma}_2^2$ of $\gamma_2$ decouple from the theory, while
A-branes consisting of preimages $\wt{\gamma}_1^\mu$ of $\gamma_1$ flow to the corresponding A-branes in the IR. 
All other A-branes decay into sums of A-branes of the two types. More precisely, 
if $\Gamma_1$ and $\Gamma_2$ intersect in $b_{a_2^1}=b_{a_1^\nu}$, then A-branes $\overline{b_{a_2^2}x^*b_{a_1^\mu}}$ in the UV decay into sums $\overline{b_{a_2^2}x^*b_{a_2^1}}+\overline{b_{a_1^\nu}x^*b_{a_1^\mu}}$ whose first summand
decouples in the IR, while the second one stays in the theory. For the case $d=4$ this is schematically represented in Figure
\ref{pertfig}. Which of the rays $\overline{x^*b_a}$ is torn off the UV critical point, and whether $b_a$ is connected to $b_{a-1}$ or $b_{a+1}$ by $\Gamma_2$ depends on the phase of the perturbation parameter $\lambda$. 

More generally, the topology of the graphical representation of a deformation depends on the form of $W_\lambda$
in a complicated way. Since the graphical representation
carries the information relevant for the analysis of the behaviour of A-branes under the respective flows, we will avoid working directly with the deformations $W_\lambda$ of the superpotential in the following, but instead characterise a perturbation directly by the graphical  representation.
\subsection{Comparison}
Indeed, the graphical representation of the behaviour of A-branes under bulk flows in the mirror LG models described in the previous section is very reminiscent of the operation of the defects $P^{(m,n)}$ on B-branes in the corresponding LG orbifolds. 
In Section \ref{picture} above, we gave a pictorial representation of B-branes in the LG-orbifolds $X^d/\ZZ_d$, in which the B-brane associated to a matrix factorisation $Q^{(M,N)}$ was represented by a union of consecutive segments $S_M\cup\ldots\cup S_{M+N-1}$ of a disk divided into $d$ segments. It can be easily worked out that the graphical representation of the corresponding
mirror A-brane in the unorbifolded LG model with superpotential $W=X^d$ is given by $\overline{b_Mx^*b_{M+N}}$. Thus, the mirror map just replaces a union of consecutive segments by its oriented boundary. 

\FIGURE[r]{
\includegraphics[width=50mm]{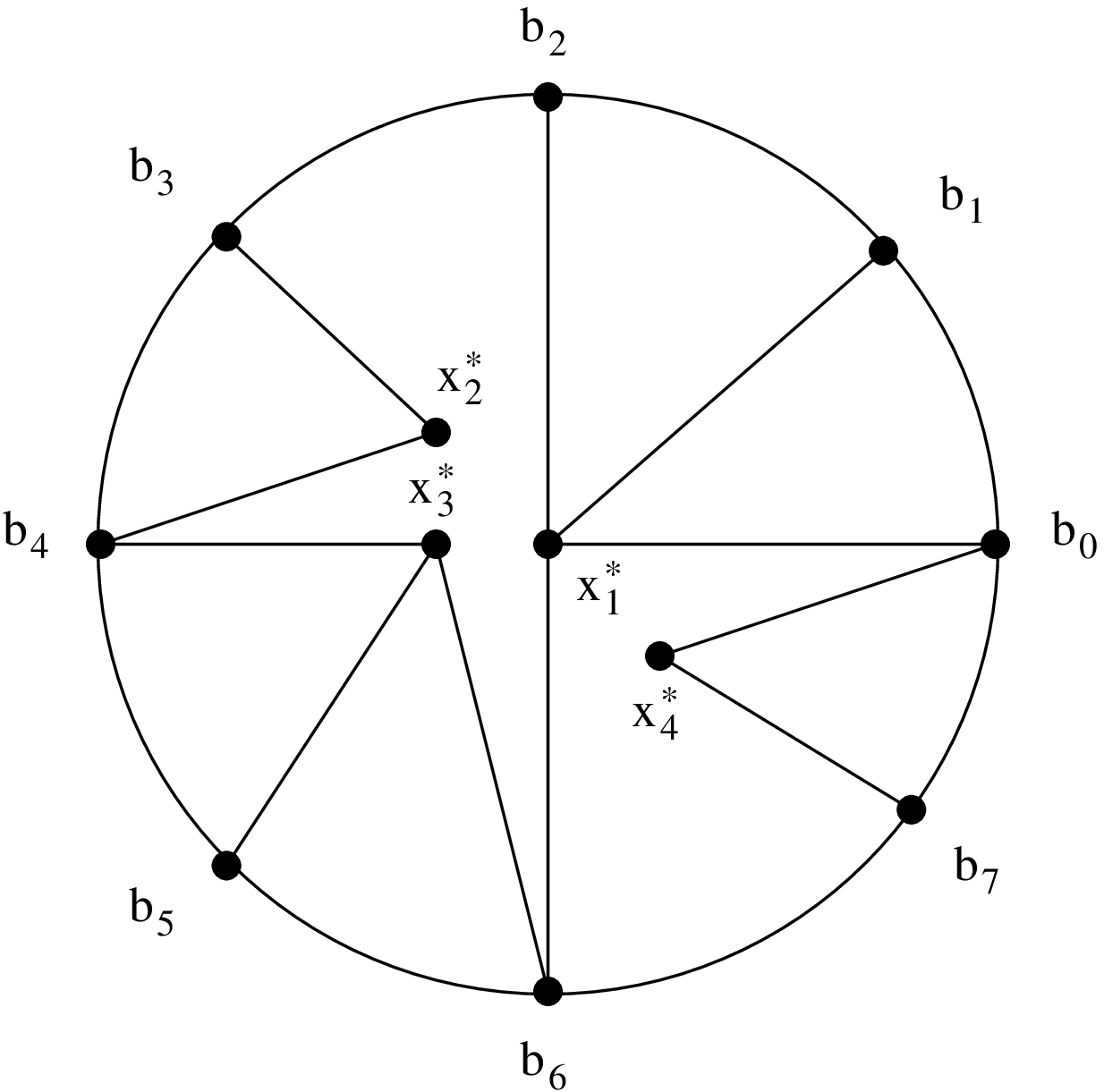}
\caption{Representation of a deformation of a degree $8$ polynomial with four critical points.
\label{diskfig}
}}
It is now obvious that under the mirror map bulk flows of
A-branes encoded in graphical representation $\{\Gamma_i\}$ such 
as in Figure \ref{diskfig} can be pictorially represented by shrinking sectors, and can therefore be described by defects $P^{(m,n)}$. More precisely, let
\beq
{\cal L}=\{a\in\ZZ_d\,|\,b_a\nsim b_{a+1}\}\,,
\eeq
be the set of neighbouring points $b_a$ which are not connected on $\Gamma-\Gamma_1$, 
and denote the complement by ${\cal L}^c$. Then the corresponding 
flow on the B-side can be represented pictorially by the shrinking operation
${\cal S}_{{\cal L}^c}$ defined in Section \ref{picture},
and thus it is realised by the corresponding defect $P^{(m,n)}$ with
\beq
{\cal L}=m+\{0,n_1,n_1+n_2,\ldots,n_1+\ldots+n_{o(x_1^*)+1}\}\,.
\eeq
Note that this parametrisation of ${\cal L}$ is ambiguous. Namely one can shift $m\mapsto m+\sum_{i=1}^j n_i$ and
change the $n_i$ accordingly. This operation is nothing else than the IR quantum symmetry, which we identified above as
giving rise to an ambiguity of the flow on A-branes.

For instance, the perturbation corresponding to Figure \ref{diskfig} can be represented on the B-side 
by the shrinking operation
${\cal S}_{\{3,4,5,7\}}$ and is therefore described by the defect $P^{(0,(1,1,4,2))}$. 

To summarise, the analysis of the induced flows of A-branes 
the mirror Landau-Ginzburg models indeed confirms
that the defects $P^{(m,n)}$ describe the flows between minimal model orbifolds.
\section{Flows between $\CC/\ZZ_d$-Orbifolds}\label{orbsec}
There is a close link between Landau-Ginzburg models with superpotential
$W$ and non-compact affine orbifold theories that can be obtained from the former
by letting the superpotential go to zero.
Although the theories are in fact quite different, for example have different
central charge and F-terms, the structure of their twisted chiral sectors
(twisted F-terms) is unaffected by the presence or absence 
of an untwisted chiral superpotential \cite{Hori:2000kt}.

Thus, the discussion of twisted chiral perturbations of Landau-Ginzburg orbifolds above
carries over to the case of affine orbifold models of type $\CC/\ZZ_d$, which can be regarded
as $\ZZ_d$ orbifolds of Landau-Ginzburg models of a single chiral superfield with superpotential $W=0$.
Indeed, these models have the same twisted chiral rings
as the minimal model orbifolds ${\cal M}_{d-2}/\ZZ_d$ with one $(a,c)$ field coming from 
the ground state of each twisted sector.
The perturbations we have been studying for the minimal model
are hence directly related to the perturbations by twisted chiral
fields in the affine orbifold models. Indeed, it has been found in \cite{Harvey:2001wm,Adams:2001sv}
that non-supersymmetric orbifold singularities of type $\CC/\ZZ_d$
flow under perturbation by twisted chiral fields into a number of
disconnected lower orbifold singularities in the IR, which is analogous to what
one finds for the minimal model orbifolds.
In fact, there is a common treatment of the corresponding flows 
${\cal M}_{d-2}/\ZZ_d\mapsto{\cal M}_{d\p-2}/\ZZ_{d\p-2}$ between 
minimal model orbifolds and 
$\CC/\ZZ_d\mapsto\CC/\ZZ_{d\p}$ between affine orbifolds
in the framework of gauged linear sigma models (see \eg \cite{Vafa:2001ra}).
This suggests that also the flows between affine orbifolds 
can be described by defects with a structure similar to that of the 
$P^{(m,n)}$.

Generally, matrix factorisations $P$ of any polynomial $W$ give rise to matrix factorisations of 
$W=0$ by setting $p_0=0$. In this way, from the matrix factorisations $P^{(m,n)}$ defining defects between LG orbifolds
$X^d/\ZZ_d$ one also obtains defects between orbifolds $\CC/\ZZ_d$. Obviously they
obey the same fusion algebra as the ones in the LG orbifolds, and also their action on
B-type defects is similar. Therefore these defects are the natural candidates 
to describe the corresponding flows between the affine orbifolds.

As a side remark we would like to mention that for Landau-Ginzburg (orbifold) 
models with superpotential $W=0$, fusion with a defect corresponding to a matrix factorisation $P=(p_1,0)$
can also be thought of as Fourier-Mukai transform with kernel the (equivariant) sheaf associated to the module 
$\coker(p_1)$.
This description is more in line with the common description of D-branes in these models
in terms of (equivariant) coherent sheaves. 
\section{Discussion}\label{discussion}
In this paper, we have considered the behaviour of B-type D-branes in $\ZZ_d$-orbifolds
of $N=2$ minimal models ${\cal M}_{d-2}$ under bulk perturbations generated by relevant twisted chiral operators.
The new approach put forward here is based on the idea that perturbations of conformal field
theories give rise to defect lines between the UV and the IR theory of the corresponding renormalisation group flows.
This turns out to be particularly useful in the treatment of bulk perturbations on surfaces with boundaries. Namely, 
the effect of bulk flows on the boundary conditions can then be realised by merging this defect with the
respective boundary condition of the UV theory to obtain a new boundary
condition of the IR theory. A related idea has been put forward in 
\cite{Bachas:2004sy}, where it was shown how certain boundary RG flows can be universally induced
by fusion with defects.
In situations where $N=2$ supersymmetry is preserved, 
the fusion procedure can be performed on the level of the respective topologically twisted theories, 
making it unnecessary to deal with regularisation issues.

Using the Landau-Ginzburg representation, we constructed a set 
of B-type defects between minimal model orbifolds as equivariant matrix factorisations of the difference of the respective
superpotentials, and we proposed them to be associated to bulk flows between these models. Their fusion among themselves
and with B-type boundary conditions was easily computed in the matrix factorisation formalism,
and a comparison with the chiral perturbations of the mirror LG models confirms that the defects indeed 
have the correct properties to describe the flows.

We also argued that in an analogous way 
one can construct similar defects which describe corresponding 
flows between affine orbifolds of type $\CC/\ZZ_d$.

Having obtained defects arising in flows between Landau-Ginzburg (or affine) orbifolds with a single
chiral superfield, it would be very interesting to find defects describing flows between such models
with several variables. In these models the analysis of the behaviour of A-branes under the corresponding flow in the mirror theories 
is much more complicated, so that the defect approach would be very useful.
It would allow the explicit computation of flows of B-branes under bulk perturbations
for instance in the $\CC^2/\Gamma$-orbifolds studied in
\cite{Harvey:2001wm,Martinec:2002wg,Moore:2004yt,Moore:2005wp}.

Besides these special examples, we expect our approach to  be powerful in any
situations where world sheet supersymmetry (as opposed to space-time
supersymmetry) is preserved. The extension to non-supersymmetric theories,
or an understanding of our flow defects on the level of the full conformal
field theory as opposed to its topological subsector,
is less straight forward, because it requires a regularisation procedure for the fusion of non-topological defects with boundary conditions. The 
investigation of the fusion properties of non-topological defects 
on the level of the full conformal field theory has recently been started
in \cite{Bachas} for the example of the free boson. One of the conclusions of that
paper was that non-topological defects are generically unstable and
tend to decay 
via defect-dissociation, the inverse process of fusion. The defects investigated in the current paper are certainly
non-topological on the level of the full conformal field theory, and one might
wonder what possible decay channels could arise. A part of the answer 
is already given in
section \ref{flowdefects}, where we have shown that defects between the minimal models
${\cal M}_{d+n-2}$ and ${\cal M}_{d-2}$ can be obtained by fusing $n$ single step
defects that relate ${\cal M}_{d+i-1}$ and ${\cal M}_{d+i-2}$. It would be an
interesting problem to determine via an analysis of the defect entropy proposed
in \cite{Bachas} wether our defects tend to dissociate into smaller step operators.

While we focused on relevant perturbations in this paper, by the same reasoning defects can also be used in
the study of exactly marginal bulk perturbations. (These do not necessarily
stay marginal in the presence of boundaries \cite{Fredenhagen:2006dn} but can induce non-trivial
RG flows in the boundary sectors.)
For instance, $\sigma$-models on Calabi-Yau target spaces have no tachyons, and hence do not exhibit relevant perturbations.
But they do allow for exactly marginal perturbations in general. Deforming such a theory around a singularity in its K\"ahler moduli space,
the corresponding monodromy transformation on the B-type D-branes should be described by a defect.
Indeed this is not at all surprising, because 
these transformation can be represented as Fourier-Mukai transformations (see \eg \cite{Aspinwall:2004jr}), which at least on
a superficial level are related to defects via the folding trick. 
The defect representation of these transformations in the Landau-Ginzburg phase is formulated
in \cite{Jockers}, see also \cite{Jockers:2006sm}.

Even though there are no relevant flows between different Calabi-Yau $\sigma$-models,
it is still possible to construct defects between such models, 
for instance using the ones constructed
in this paper for single minimal model orbifolds as building blocks.
The physical meaning of such defects however is unclear.
Clearly they relate different string vacua, and one might speculate that defect transitions could require some meaning, \eg as tunneling amplitudes in a background independent formulation of string theory. 
\section*{Acknowledgements}
DR would like to thank ETH Z\"urich for hospitality, where part of this work was done. We also thank the Centre Interfacultaire de Bernoulli and in particular
the organisers of the program on matrix factorisations for a stimulating
workshop. We would like to thank C.~Bachas, M.~Gaberdiel, M.~Herbst, K.~Hori, W.~Lerche for
discussions and correspondence.
The work of IB is supported by an EURYI award and the Marie-Curie network
Forces Universe (MRTN-CT-2004-005104). DR was supported by a DFG research fellowship and partially by DOE grant DE-FG02-96ER40959.
\appendix
\section{The Conformal Field Theory Point of View}\label{cftapp}
In this appendix we would like to discuss various features
that appeared in the main text from the point of view of the
conformal field theory, to which the LG model flows in the IR.
Our discussion will be restricted to defects that preserve
the superconformal symmetry, in particular, we will only consider
defects between one minimal model and itself. 

In the IR, the Landau-Ginzburg model with one chiral superfield
and superpotential $W(X)=X^d$ 
flows to 
the unitary superconformal minimal
model ${\mathcal M}_k$, $k=d-2$ with A-type modular invariant
partition function. 
These conformal field theories are rational with respect to the $N=2$ super
Virasoro algebra at central charge $c_k={3k\over k+2}$. In fact, the
bosonic part of this algebra can be realised as the coset W-algebra
\beq\label{coset}
\left({\rm SVir}_{c_k}\right)_{\rm bos} = 
{\widehat{\mathfrak{su}}(2)_k\oplus\widehat{\mathfrak{u}}(1)_4\over \widehat{\mathfrak{u}}(1)_{2k+4}}\,,
\eeq  
and the respective coset CFT can be obtained from ${\mathcal M}_k$ by
a non-chiral GSO projection. 

The Hilbert space $\HH^k$ of ${\mathcal M}_k$ decomposes into irreducible
highest weight representations of holomorphic and antiholomorphic
super Virasoro algebras, but it is convenient to decompose it further
into irreducible highest weight representations $\VV_{[l,m,s]}$ of the
bosonic subalgebra \eq{coset}. These representations are labelled by 
\beq
[l,m,s]\in\mathcal{I}_k:=\{(l,m,s)\,|\,0\leq l\leq
k,\,m\in\ZZ_{2k+4},\,s\in\ZZ_4,\,l+m+s\in 2\ZZ\}/\sim\,,
\eeq
where $[l,m,s]\sim[k-l,m+k+2,s+2]$ is the field identification. The highest weight representations
of the full super Virasoro algebra are given by
\beq
\VV_{[l,m]}:=\VV_{[l,m,(l+m)\,{\rm
    mod}\,2]}\oplus\VV_{[l,m,(l+m)\,{\rm mod}\,2+2]}\,.
\eeq
For $(l+m)$ even $\VV_{[l,m]}$ is in the NS-, for $(l+m)$ odd in the R-sector.
Here $[l,m]\in{\mathcal J}_k:=\{(l,m)\,|\,0\leq l\leq
k,\,m\in\ZZ_{2k+4}\}/\sim$, $[l,m]\sim[k-l,m+k+2]$. 
The Hilbert spaces of ${\mathcal
M}_k$ in the NSNS- and RR-sectors then read
\beq
\HH^k_{NSNS} 
\cong\bigoplus_{\stackrel{[l,m]\in{\mathcal J}_k}{l+m\,{\rm
      even}}}\VV_{[l,m]}\otimes\overline{\VV}_{[l,m]} \,,\qquad
\HH^k_{RR}
\cong\bigoplus_{\stackrel{[l,m]\in{\mathcal J}_k}{l+m\,{\rm
      odd}}}\VV_{[l,m]}\otimes\overline{\VV}_{[l,m]} \,.
\eeq
The theory exhibits an action of a $\ZZ_{k+2}$ symmetry group, realised by
the simple current $(0,2,0)$. Orbifolding by this group introduces
twisted sectors, in which the representations of
the left- and right-movers differ by  the
action of the appropriate power of the simple current. Having included the twisted sectors,
one has to projects onto $\ZZ_{k+2}$-invariant sectors to obtain the Hilbert space of the orbifold theory.
The action of the generator of the orbifold group in the twisted sector $\psi \in \VV_{[l,m,s]} \otimes \VV_{[l,m-2n,s]}$
is given by multiplication with the phase
\beq\label{orbproject}
\psi\mapsto e^{2\pi i \frac{m+m-2n}{2(k+2)}}
\psi \,.
\eeq
The resulting Hilbert space differs from the initial unorbifolded one only by a relative minus
sign of $m$-labels in the left- and right-moving sectors:
\beq
\HH^k_{NSNS} 
\cong\bigoplus_{\stackrel{[l,m]\in{\mathcal J}_k}{l+m\,{\rm
      even}}}\VV_{[l,m]}\otimes\overline{\VV}_{[l,-m]} \,,\qquad
\HH^k_{RR}
\cong\bigoplus_{\stackrel{[l,m]\in{\mathcal J}_k}{l+m\,{\rm
      odd}}}\VV_{[l,m]}\otimes\overline{\VV}_{[l,-m]} \,.
\eeq
\subsection{Defects}
B-type defects in minimal models have been considered in 
\cite{Brunner:2007qu}. Here,
defects were formulated as maps between closed string Hilbert spaces.
They can be written as sums over projectors onto modules of the
bosonic subalgebra of the full supersymmetric model: 
\beq\label{unorbdefop}
{\mathcal D}=\sum_{\stackrel{[l,m,s],\bar s}{s-\bar s\,{\rm
      even}}}{\mathcal D}^{[l,m,s,\bar s]} {\rm P}_{[l,m,s,\bar s]}\,,
\eeq
where ${\rm P}_{[l,m,s,\bar s]}$ is a projector on the subspace $
\VV_{[l,m,s]} \otimes \VV_{[l,m,\bar{s}]}$ of the Hilbert space.
It is furthermore understood that 
\beq
{\mathcal
  D}^{[l,m,s+2,\bar s]}=\eta{\mathcal D}^{[l,m,s,\bar
  s]}\quad{\rm and}\quad 
{\mathcal
  D}^{[l,m,s,\bar s+2]}=\bar \eta{\mathcal D}^{[l,m,s,\bar
  s]}\,.
\eeq
Consistent choices for the prefactors of the projection operators
are given by
\beq\label{qdim}
{\mathcal D}_{[ L, M,
S,\bar S]}^{[l,m,s,\bar s]}=e^{-i\pi {\bar S(s+\bar s)\over 2}}{S_{[ L, M,
  S-\bar S][l,m,s]}\over
S_{[0,0,0],[l,m,s]}}\,,
\eeq
where the different defects have been labelled by $[ L,
M, S,\bar S]$ with $[L,M,S-\bar S]\in {\mathcal I}_k$, and 
\beq
S_{[L,M,S][l,m,s]}={1\over k+2}e^{-i\pi{Ss\over 2}}e^{i\pi{Mm\over
k+2}}\sin\left(\pi{(L+1)(l+1)\over k+2}\right)
\eeq
is the modular $S$-matrix for the coset representations
$\VV_{[l,m,s]}$. 
It is then straightforward to determine the composition of defects
and their action on boundary states \cite{Brunner:2007qu}.

To obtain the defect in the orbifold theory, one simply has to
switch the sign of $m$ for the right movers, such that the
defect reads
\beq\label{orbdefop}
{\mathcal D}^{\rm orb}=\sum_{\stackrel{[l,m,s],\bar s}{s-\bar s\,{\rm
      even}}}{\mathcal D}^{[l,m,s,\bar s]} {\rm P}^-_{[l,m,s,\bar s]}\,,
\eeq
where ${\rm P}^-_{[l,m,s,\bar s]}$ is a projector on the subspace $
\VV_{[l,m,s]} \otimes \VV_{[l,-m,\bar{s}]}$ of the Hilbert space.

\subsection{The Folding Trick}

The folding trick relates defects to permutation boundary
states \cite{Recknagel:2002qq} of the tensor product of two minimal models. We start with
the unorbifolded theory. 
B-type permutation boundary states in a tensor product of two minimal models
satisfy the following conditions
\beqa
\big( G_{r}^{\pm (1)} + i\eta_1 \bar{G}_{-r}^{\pm (2)} \big) \kket{B} &=& 0 \\
\nn
\big( G_{r}^{\pm (2)} + i\eta_2 \bar{G}_{-r}^{\pm (1)} \big) \kket{B} &=& 0
\eeqa
In the case $\eta_1=\eta_2$ the boundary conditions preserve the diagonal
$N=2$ algebra. The corresponding boundary states have been discussed in
\cite{Brunner:2005fv,Enger:2005jk} and are explicitely given by
\beq\label{bpermstate}
\kket{[L,M,S_1,S_2]} = \frac{1}{2\sqrt{2}} \!\sum_{l,m,s_1,s_2}\!
\frac{S_{Ll}}{S_{0l}} e^{\pi i Mm/(k+2)} e^{-i\pi (S_1s_1-S_2s_2)/2}
\iket{[l,m,s_1]\otimes [l,-m, -s_2]}
\eeq
Permutation boundary states in minimal model orbifolds can now be constructed using
standard conformal field theory techniques.
We first note that the B-type permutation boundary states \eq{bpermstate} in the unorbifolded theory
are invariant
under the diagonal subgroup $\ZZ_d \subset \ZZ_d \times \ZZ_d$ generated
by the product $g=g_1 g_2$ of the generators of the two $\ZZ_d$'s. 
To construct the $g^n$-twisted components of the boundary states
we observe that the permutation gluing condition requires that 
$\bar{m}_2 = -m_1$ and $m_2 = - \bar{m}_1$. In the sector twisted by
$g^n$ the relation between left- and right-moving $m$-labels is
$m_1=\bar{m}_1 +2n$, $m_2=\bar{m}_2 +2n$, so that the relevant 
Ishibashi states have labels $m_2=-\bar{m}_1 = -m_1+2n$. Therefore,
the twisted boundary states take the form
\begin{eqnarray}
|\!|L,M,\hat{M},S_1,S_2\rangle\!\rangle_{(-1)^{(s+1)F}g^n} & = & 
\frac{1}{2} e^{-\frac{\pi i n}{k+2}(M+\hat{M})}
\sum_{l,m} \sum_{\nu_,\nu_2\in \ZZ_2}  
\frac{S_{Ll}}{S_{0l}}\, e^{\pi i \frac{Mm}{k+2}} \,
(-1)^{S_1 \nu_1+S_2\nu_2} \nonumber \\
&& \; e^{-\pi i \frac{s}{2}(S_1+S_2)} \, 
|[l,m,s+2\nu_1]\otimes [l,-m+2n ,s+2\nu_2]\rangle\!\rangle
\,, \nonumber 
\end{eqnarray}
where the additional label $\hat{M}$ specifies the 
representation of the diagonal $\ZZ_d$ on the Chan-Paton factors.
The subscript denotes the twist: for $g^n$ the Ishibashi
states are in the $n^{th}$ twisted sector. Furthermore, $s$ distinguishes
between NS and R sector, in our notation the NS sector is the $(-1)^F$
twisted R-sector.
We require that $M+\hat{M}$ is always even, so that the boundary state
is invariant under $n\to n+k+2$. Also, as before, to preserve the
diagonal $N=2$ we require that
$L+M$ and $S_1+S_2$ are even.

To obtain a boundary state that is invariant under the full $\ZZ_d \times
\ZZ_d $ orbifold group, we need to perform the projection (\ref{orbproject})
on states with $2m=2n$ mod $2k+4$.
This yields the following boundary state
\begin{eqnarray}
|\!|L,\hat{M},S_1,S_2\rangle\!\rangle_{(-1)^{(s+1)F}} & = & 
\frac{1}{2} 
\sum_{l,m} \sum_{\nu_,\nu_2\in \ZZ_2}  
\frac{S_{Ll}}{S_{0l}}\, e^{\pi i \frac{\hat{M}m}{k+2}} \,
(-1)^{S_1 \nu_1+S_2\nu_2} \nonumber \\
&& \; e^{-\pi i \frac{s}{2}(S_1+S_2)} \, 
|[l,m,s+2\nu_1]\otimes [l,m ,s+2\nu_2]\rangle\!\rangle
\, . \nonumber 
\end{eqnarray}
In this way we have constructed B-type permutation boundary states in the tensor product of minimal model
orbifolds out of those in the corresponding unorbifolded theory. This orbifold procedure is 
analogous to the one described on the level of Landau-Ginzburg models in Section \ref{deflgorb}.
In particular, after unfolding the
states with $L=0$ correspond to the defects realising the group of quantum symmetries
in the 
orbifold theory which have been constructed in Landau Ginzburg formalism in Section \ref{qsdef}.

\subsection{Cylinder Amplitude and the Folding Trick}

From the formula of the defect operators \ref{orbdefop}, it is straighforward to
determine the fusion of the corresponding defects with D-branes.
Instead of doing this calculation, we find it instructive 
to present an alternative derivation using the folding trick.
More specifically,
we will compute cylinder amplitudes in the tensor product of minimal models between permutation boundary 
states on one side and tensor product boundary states on the other. Via the folding trick we will reinterprete
them as cylinder amplitudes in a single minimal model with boundary conditions corresponding to the two tensor factors
on both ends of the cylinder with a defect line corresponding to the permutation boundary state in between them. 
The relevant one-loop
amplitude in the unorbifolded theory is
\bea
&& \langle\!\langle  (L_1,S_1)|\!|\otimes\langle\!\langle(L_2,S_2) |\!| 
q^{\frac{1}{2}(L_0 + \bar{L}_0) - \frac{c}{12}}
\,\, |\!| \,\, [\hat{L},\hat{M},\hat{S}_1,\hat{S}_2] \rangle\!\rangle   
\label{relov} \\
& & \qquad =  \sum_{[l,m,s]}  
\chi_{[l,m,s]}(\tilde{q}^{1/2}) \,
\sum_{\hat{l}} 
\Bigl( N_{L_1 L_2}{}^{\hat{l}}\, N_{\hat{l} \hat{L}}{}^{l} \,
\delta^{(4)}(s+\hat{S}_1+\hat{S}_2 - (S_1+S_2) + 1 ) \nonumber \\
& & \qquad
\qquad \qquad \qquad \qquad\qquad \quad
+ N_{k- L_1 L_2}{}^{\hat{l}}\, N_{\hat{l} \hat{L}}{}^{l} \,
\delta^{(4)}(s+\hat{S}_1+\hat{S}_2 - (S_1+S_2) - 1 ) \Bigr)\,. \nonumber 
\eea
Here $|\!|(L_i,S_i)\rangle\!\rangle$ are B-type boundary state 
in a single minimal model (see e.g. \cite{Brunner:2007qu} for more
details on the notation).
Note that in the case $\hat{S_1}=\hat{S_2}$ mod $2$  and $S_1=S_2$ mod $2$
the representations appearing in the open string sector are formally
in the R-sector, but are to be interpreted as twisted NS-sector 
representations. In the closed string sector, this shift is related
to the fact that the overlap of a tensor product with a permutation boundary
state is a trace with an insertion of the permutation $\sigma$. Since $\sigma$
interchanges states, a minus sign is picked up in the fermionic relative
to the bosonic case.
\beqn
\langle\!\langle [l,0,s]\otimes [l,0,s] | 
q^{\frac{1}{2}(L_0+ \bar{L}_0)-\frac{c}{12}} 
| [l,0,s]\otimes [l,0,s] \rangle\!\rangle^\sigma &=&
{\rm Tr}_{[l,0,s]\otimes [l,0,s]} \nn
\left(q^{L_0-\frac{c}{12}} \,\sigma\right)   \\ 
&=& e^{-\pi i s/2} \chi_{[l,0,s]} (q^2) \ .
\eeqn
We would now like to find the defect ${\mathcal D}$
corresponding to the permutation boundary
state. This can be deduced by comparing the characters appearing in this 
cylinder amplitude with those appearing in the cylinder amplitudes
between two B-type boundary states 
in a single minimal model. The goal is to find  a homomorphism ${\mathcal D}$ such
that the above amplitude is reproduced by the cylinder amplitude between ${\mathcal D}
\kket{(L_1,S_1)}$ on one side and $\kket{(L_2,-S_2)}$ on the other.
Note that in the cylinder amplitude
taken in the tensor product, the boundary states $\kket{(L_1,S_1)}$ and $\kket{(L_2,S_2)}$
are both ingoing (or both outgoing). 
On the other hand, taking a cylinder amplitude in a
single minimal model, one of the boundaries becomes outgoing (ingoing), so
that one of the states has to be conjugated: $\kket{(L,S)}\mapsto\langle\!\langle(L,-S)|\!|$.
Further care must be taken because of the phase $(-1)^{F_L}$ that appears
in the folded model. Taking this phase into account effectively
shifts the $S$-label of a B-type boundary state by one (changing the spin
structure), such that the $\delta^{(4)}$-constraint in the above formula
gets shifted by one.
Taking all of this into account, the above formula is consistent with
the defect action
\beqn\label{CFTdefoperation}
{\mathcal D}_{[ L_1, M_1, S_1,\bar S_1]}\kket{[ L_2, M_2, S_2]}_B
&=&\!\!\!\sum_{[ L, M, S]\in{\mathcal I}_k}\!\!\!{\mathcal
N}_{[ L_1, M_1, S_1-\bar S_1][ L_2,M_2, S_2]}^{[
L, M, S]}\kket{[ L, M, S]}_B\\
&=&\sum_{ L}{\mathcal N}_{ L_1 L_2}^{ L} \kket{[
L, M_1+ M_2, S_1-\bar S_1+ S_2]}_B\, ,\nonumber
\eeqn
such that the permutation boundary state corresponds to the defect operator \eq{unorbdefop}
with the same labels. The discussion in the orbifold theory is similar, the only difference
being that the boundary states of the orbifold have an additional $M$-label,
leading to a $\delta^{(2k+4)}$ constraint on the $m$-labels in all cylinder
amplitudes.
$$ $$
\bibliographystyle{flowdef}
\bibliography{flowdef}

\providecommand{\href}[2]{#2}\begingroup\raggedright\begin{thebibliography}{10}

\bibitem{Adams:2001sv}
A.~Adams, J.~Polchinski, and E.~Silverstein, ``Don't panic! {C}losed string
  tachyons in {ALE} space-times,'' {\em JHEP} {\bf 10} (2001) 029,
\href{http://arXiv.org/abs/hep-th/0108075}{{\tt hep-th/0108075}}.

\bibitem{Harvey:2001wm}
J.~A. Harvey, D.~Kutasov, E.~J. Martinec, and G.~W. Moore, ``Localized tachyons
  and {RG} flows,''
\href{http://arXiv.org/abs/hep-th/0111154}{{\tt hep-th/0111154}}.

\bibitem{Vafa:2001ra}
C.~Vafa, ``Mirror symmetry and closed string tachyon condensation,''
\href{http://arXiv.org/abs/hep-th/0111051}{{\tt hep-th/0111051}}.

\bibitem{Martinec:2002wg}
E.~J. Martinec and G.~W. Moore, ``On decay of {K}-theory,''
\href{http://arXiv.org/abs/hep-th/0212059}{{\tt hep-th/0212059}}.

\bibitem{Moore:2004yt}
G.~W. Moore and A.~Parnachev, ``Localized tachyons and the quantum {McKay}
  correspondence,'' {\em JHEP} {\bf 11} (2004) 086,
\href{http://arXiv.org/abs/hep-th/0403016}{{\tt hep-th/0403016}}.

\bibitem{Moore:2005wp}
G.~Moore and A.~Parnachev, ``Profiling the brane drain in a nonsupersymmetric
  orbifold,'' {\em JHEP} {\bf 01} (2006) 024,
\href{http://arXiv.org/abs/hep-th/0507190}{{\tt hep-th/0507190}}.

\bibitem{Petkova:2000ip}
V.~B. Petkova and J.~B. Zuber, ``Generalised twisted partition functions,''
  {\em Phys. Lett.} {\bf B504} (2001) 157--164,
\href{http://arXiv.org/abs/hep-th/0011021}{{\tt hep-th/0011021}}.

\bibitem{Bachas:2001vj}
C.~Bachas, J.~de~Boer, R.~Dijkgraaf, and H.~Ooguri, ``Permeable conformal walls
  and holography,'' {\em JHEP} {\bf 06} (2002) 027,
\href{http://arXiv.org/abs/hep-th/0111210}{{\tt hep-th/0111210}}.

\bibitem{Frohlich:2004ef}
J.~Fr{\"o}hlich, J.~Fuchs, I.~Runkel, and C.~Schweigert, ``Kramers-{W}annier
  duality from conformal defects,'' {\em Phys. Rev. Lett.} {\bf 93} (2004)
  070601,
\href{http://arXiv.org/abs/cond-mat/0404051}{{\tt cond-mat/0404051}}.

\bibitem{Bachas:2004sy}
C.~Bachas and M.~Gaberdiel, ``Loop operators and the {Kondo} problem,'' {\em
  JHEP} {\bf 11} (2004) 065,
\href{http://arXiv.org/abs/hep-th/0411067}{{\tt hep-th/0411067}}.

\bibitem{Frohlich:2006ch}
J.~Fr{\"o}hlich, J.~Fuchs, I.~Runkel, and C.~Schweigert, ``Duality and defects
  in rational conformal field theory,'' {\em Nucl. Phys.} {\bf B763} (2007)
  354--430,
\href{http://arXiv.org/abs/hep-th/0607247}{{\tt hep-th/0607247}}.

\bibitem{Quella:2006de}
T.~Quella, I.~Runkel, and G.~M.~T. Watts, ``Reflection and transmission for
  conformal defects,'' {\em JHEP} {\bf 04} (2007) 095,
\href{http://arXiv.org/abs/hep-th/0611296}{{\tt hep-th/0611296}}.

\bibitem{Alekseev:2007in}
A.~Alekseev and S.~Monnier, ``Quantization of {Wilson} loops in
  {Wess}-{Zumino}-{Witten} models,'' {\em JHEP} {\bf 08} (2007) 039,
\href{http://arXiv.org/abs/hep-th/0702174}{{\tt hep-th/0702174}}.

\bibitem{Fuchs:2007tx}
J.~Fuchs, M.~R. Gaberdiel, I.~Runkel, and C.~Schweigert, ``Topological defects
  for the free boson {CFT},''
\href{http://arXiv.org/abs/arXiv:0705.3129 [hep-th]}{{\tt arXiv:0705.3129
  [hep-th]}}.

\bibitem{Runkel:2007wd}
I.~Runkel, ``Perturbed defects and {T}-systems in conformal field theory,''
\href{http://arXiv.org/abs/arXiv:0711.0102 [hep-th]}{{\tt arXiv:0711.0102
  [hep-th]}}.

\bibitem{Khovanov:2004bc}
M.~Khovanov and L.~Rozansky, ``Topological {L}andau-{G}inzburg models on a
  world-sheet foam,''
\href{http://arXiv.org/abs/hep-th/0404189}{{\tt hep-th/0404189}}.

\bibitem{Kapustin:2004df}
A.~Kapustin and L.~Rozansky, ``On the relation between open and closed
  topological strings,'' {\em Commun. Math. Phys.} {\bf 252} (2004) 393--414,
\href{http://arXiv.org/abs/hep-th/0405232}{{\tt hep-th/0405232}}.

\bibitem{Brunner:2007qu}
I.~Brunner and D.~Roggenkamp, ``{B}-type defects in {L}andau-{G}inzburg
  models,'' {\em JHEP} {\bf 08} (2007) 093,
\href{http://arXiv.org/abs/arXiv:0707.0922 [hep-th]}{{\tt arXiv:0707.0922
  [hep-th]}}.

\bibitem{Dorey:1997yg}
P.~Dorey, A.~Pocklington, R.~Tateo, and G.~Watts, ``{TBA} and {TCSA} with
  boundaries and excited states,'' {\em Nucl. Phys.} {\bf B525} (1998)
  641--663,
\href{http://arXiv.org/abs/hep-th/9712197}{{\tt hep-th/9712197}}.

\bibitem{Dorey:2000eh}
P.~Dorey, M.~Pillin, R.~Tateo, and G.~M.~T. Watts, ``One-point functions in
  perturbed boundary conformal field theories,'' {\em Nucl. Phys.} {\bf B594}
  (2001) 625--659,
\href{http://arXiv.org/abs/hep-th/0007077}{{\tt hep-th/0007077}}.

\bibitem{Fredenhagen:2006dn}
S.~Fredenhagen, M.~R. Gaberdiel, and C.~A. Keller, ``Bulk induced boundary
  perturbations,'' {\em J. Phys.} {\bf A40} (2007) F17,
\href{http://arXiv.org/abs/hep-th/0609034}{{\tt hep-th/0609034}}.

\bibitem{Keller:2007nd}
C.~A. Keller, ``Brane backreactions and the {Fischler}-{Susskind} mechanism in
  conformal field theory,''
\href{http://arXiv.org/abs/arXiv:0709.1076 [hep-th]}{{\tt arXiv:0709.1076
  [hep-th]}}.

\bibitem{Baumgartl:2007an}
M.~Baumgartl, I.~Brunner, and M.~R. Gaberdiel, ``D-brane superpotentials and
  {RG} flows on the quintic,'' {\em JHEP} {\bf 07} (2007) 061,
\href{http://arXiv.org/abs/arXiv:0704.2666 [hep-th]}{{\tt arXiv:0704.2666
  [hep-th]}}.

\bibitem{Green:2007wr}
D.~Green, M.~Mulligan, and D.~Starr, ``Boundary entropy can increase under bulk
  {RG} flow,''
\href{http://arXiv.org/abs/arXiv:0710.4348 [hep-th]}{{\tt arXiv:0710.4348
  [hep-th]}}.

\bibitem{Bachas}
C.~Bachas and I.~Brunner, ``Fusion of conformal interfaces,''
  \href{http://arXiv.org/abs/arXiv:07012.0076}{{\tt arXiv:07012.0076}}.

\bibitem{Hori:2000ic}
K.~Hori, ``Linear models of supersymmetric {D}-branes,'' {\em Symplectic
  geometry and mirror symmetry} (Seoul 2000)
\href{http://arXiv.org/abs/hep-th/0012179}{{\tt hep-th/0012179}}.

\bibitem{Hori:2000ck}
K.~Hori, A.~Iqbal, and C.~Vafa, ``D-branes and mirror symmetry,''
\href{http://arXiv.org/abs/hep-th/0005247}{{\tt hep-th/0005247}}.

\bibitem{Kontsevich}
M.~Kontsevich (unpublished).

\bibitem{Kapustin:2002bi}
A.~Kapustin and Y.~Li, ``D-branes in {L}andau-{G}inzburg models and algebraic
  geometry,'' {\em JHEP} {\bf 12} (2003) 005,
\href{http://arXiv.org/abs/hep-th/0210296}{{\tt hep-th/0210296}}.

\bibitem{Orlov}
D.~Orlov, ``Triangulated categories of singularities and {D}-branes in
  {L}andau-{G}inzburg models,'' \href{http://arXiv.org/abs/math/0302304}{{\tt
  math/0302304}}.

\bibitem{Brunner:2003dc}
I.~Brunner, M.~Herbst, W.~Lerche, and B.~Scheuner, ``{Landau-Ginzburg}
  realization of open string {TFT},'' {\em JHEP} {\bf 11} (2006) 043,
\href{http://arXiv.org/abs/hep-th/0305133}{{\tt hep-th/0305133}}.

\bibitem{Ashok:2004zb}
S.~K. Ashok, E.~Dell'Aquila, and D.-E. Diaconescu, ``Fractional branes in
  {Landau-Ginzburg} orbifolds,'' {\em Adv. Theor. Math. Phys.} {\bf 8} (2004)
  461--513,
\href{http://arXiv.org/abs/hep-th/0401135}{{\tt hep-th/0401135}}.

\bibitem{Hori:2004ja}
K.~Hori and J.~Walcher, ``F-term equations near {Gepner} points,'' {\em JHEP}
  {\bf 01} (2005) 008,
\href{http://arXiv.org/abs/hep-th/0404196}{{\tt hep-th/0404196}}.

\bibitem{Kov-Roz}
M.~Khovanov and L.~Rozansky, ``Matrix factorizations and link homology,''
  \href{http://arXiv.org/abs/math.QA/0401268}{{\tt math.QA/0401268}}.

\bibitem{Gaberdiel:2007us}
M.~R. Gaberdiel and A.~Lawrence, ``Bulk perturbations of {N = 2} branes,'' {\em
  JHEP} {\bf 05} (2007) 087,
\href{http://arXiv.org/abs/hep-th/0702036}{{\tt hep-th/0702036}}.

\bibitem{Hori:2000kt}
K.~Hori and C.~Vafa, ``Mirror symmetry,''
\href{http://arXiv.org/abs/hep-th/0002222}{{\tt hep-th/0002222}}.

\bibitem{Aspinwall:2004jr}
P.~S. Aspinwall, ``D-branes on {C}alabi-{Y}au manifolds,''
\href{http://arXiv.org/abs/hep-th/0403166}{{\tt hep-th/0403166}}.

\bibitem{Jockers}
I.~Brunner, H.~Jockers, and D.~Roggenkamp, ``to appear,''.

\bibitem{Jockers:2006sm}
H.~Jockers, ``D-brane monodromies from a matrix-factorization perspective,''
  {\em JHEP} {\bf 02} (2007) 006,
\href{http://arXiv.org/abs/hep-th/0612095}{{\tt hep-th/0612095}}.

\bibitem{Recknagel:2002qq}
A.~Recknagel, ``Permutation branes,'' {\em JHEP} {\bf 04} (2003) 041,
\href{http://arXiv.org/abs/hep-th/0208119}{{\tt hep-th/0208119}}.

\bibitem{Brunner:2005fv}
I.~Brunner and M.~R. Gaberdiel, ``Matrix factorisations and permutation
  branes,'' {\em JHEP} {\bf 07} (2005) 012,
\href{http://arXiv.org/abs/hep-th/0503207}{{\tt hep-th/0503207}}.

\bibitem{Enger:2005jk}
H.~Enger, A.~Recknagel, and D.~Roggenkamp, ``Permutation branes and linear
  matrix factorisations,'' {\em JHEP} {\bf 01} (2006) 087,
\href{http://arXiv.org/abs/hep-th/0508053}{{\tt hep-th/0508053}}.

\end{thebibliography}\endgroup
\end{document}